\documentclass[prd,amsmath,floats,amssymb, floatfix,superscriptaddress,nofootinbib,onecolumn,preprint]{revtex4} 

\usepackage[plainpages=false, colorlinks=true, anchorcolor=blue, linkcolor=blue, citecolor=blue, bookmarks=false]{hyperref}
\usepackage[T1]{fontenc}
\usepackage{graphicx}% Include figure files
\usepackage{dcolumn}% Align table columns on decimal point
\usepackage{bm}% bold math
\usepackage{longtable}
\usepackage{adjustbox}

\usepackage{float}
\usepackage{booktabs}
\usepackage{array}
\usepackage[paperwidth=210mm,paperheight=297mm,centering,hmargin=2cm,vmargin=2.5cm]{geometry}
\usepackage{appendix}

\begin{document}
%\include{notations}
%\preprint{APS/123-QED}

\title{Search for GeV gamma-ray emission from PSZ G181.06+48.47 galaxy cluster  using Fermi-LAT data}% Force line breaks with \\

\author{Anuja Deshpande}
 \altaffiliation{Email:anuja1152004@gmail.com}
\affiliation{Fergusson College, Pune, Maharashtra 411004, India}
\author{Siddhant Manna}
 \altaffiliation{Email:ph22resch11006@iith.ac.in}%Lines break automatically or can be forced with \\
\author{Shantanu Desai}
 \altaffiliation{Email:shntn05@gmail.com}
\affiliation{
 Department of Physics, IIT Hyderabad Kandi, Telangana 502284,  India}


%\affiliation{
%\affiliation{
%\affiliation{%


             
\begin{abstract}
We present a search for high energy gamma rays in the energy range from 1--300~GeV   from the galaxy cluster PSZ2~G181.06+48.47  using 17.9~years of Fermi-LAT data. A binned likelihood analysis employing the 16-year source catalog reveals a significant $\gamma$-ray excess at the cluster position. Modelling the emission as a point source yields a detection with $\mathrm{TS}=19.0 (4.3 \sigma)$, a photon index of $\Gamma=3.20 \pm 0.62$, and an integrated photon flux of $(9.75\pm2.74)\times10^{-11}\,\mathrm{ph\,cm^{-2}\,s^{-1}}$, but leaves statistically significant residual emission at the cluster center. Replacing the point-source hypothesis with a RadialGaussian spatial template significantly improves the fit, with a preferred width of $\sigma=0.4^{\circ}$ and an extension significance of $\mathrm{TS}_{\rm ext}=36.8 (6.0\sigma)$. The adopted Gaussian model yields a source detection significance of $\mathrm{TS}=55.0 (7.4\sigma)$, a photon index of $\Gamma=2.62\pm0.26$, and an integrated photon flux of $(3.22\pm0.50)\times10^{-10}\,\mathrm{ph\,cm^{-2}\,s^{-1}}$, while reducing the residual emission at the cluster position to a level consistent with zero. Residual TS mapping further identifies a nearby $\gamma$-ray excess, and simultaneous modeling confirms that the extended emission associated with PSZ2~G181.06+48.47 is robust against contamination from neighboring sources. The cluster spectral energy distribution shows significant emission only in the lowest energy interval (1.0--3.13~GeV), while all higher-energy bins are consistent with upper limits, indicating a soft $\gamma$-ray spectrum with no evidence for emission above $\sim5$~GeV. These results provide strong evidence that the $\gamma$-ray emission associated with PSZ2~G181.06+48.47 is spatially extended on a scale of $\sigma\approx0.4^{\circ}$ and is more consistent with diffuse intracluster emission than with a single unresolved point source.
\end{abstract}

\maketitle
%\tableofcontents
\section{\label{sec:level1}Introduction\protect}
Galaxy clusters are the most massive virialized and gravitationally collapsed objects in the universe, making them exceptional laboratories for investigating cosmology \citep{Kravtsov2012, Allen2011, Vikhlininrev2014,Miyatake25} and fundamental physics~\citep{Bohringer2016,Desai2018, Bora2021}. They  have been detected throughout the electromagnetic spectrum  from radio \citep{Feretti2012} to hard X-rays \citep{Wik2014}. 

Until a few years ago, evidence for gamma-ray emission ($>=$ GeV energies) was still elusive, since initial searches for gamma-ray emission from galaxy clusters  using the flagship Fermi-LAT GeV gamma-ray telescope reported null results~\cite{Ackermann2010,Ackermann2014,Ackermann2016}. However, in the last few years statistically significant evidence for gamma-ray emission has been found from a number of  galaxy clusters~\cite{Manna2024,Harale2025,ShangLi,Baghmanyan2022} as well as from a stacked analysis of galaxy clusters~\cite{ReissKeshet2018,Keshet2025a,Mannastacked}. The individual clusters for which gamma-ray emission with statistical significance greater than $4.0 \sigma$ have been detected include Coma, Abell 3667, Abell 119, Abell 2065, and Abell 2244. However, it is still unclear whether this signal is coming from the intra-cluster medium, or due to point sources within the cluster or dark matter annihilation. Nevertheless, one common property of all the aforementioned clusters is that radio relics have been detected in all of them. Radio relics provide a signature of cosmic-ray electrons accelerated with Lorentz factors $> 1000$  and are produced due to shocks resulting from mergers of clusters~\cite{Brunetti14}. Gamma-ray emission could be produced due to inverse Compton scattering off these relativistic electrons or hadronic interactions of cosmic ray protons interacting with the intra-cluster medium.

In this work, we search for gamma-ray emission from PSZ2 G181.06+48.47~\cite{Ahn25,Stroe25,Rajpurohit25}. This cluster located at a redshift of $z=0.234$ is a merging cluster that contains double radio relics separated by  $\sim$ 2.7 Mpc. This cluster contains two brightest cluster galaxies with extended X-ray emission connecting the two. PSZG181 was first identified at optical wavelengths in the SDSS DR8 catalog using the redMaPPer algorithm~\cite{Rykoff} and using Sunyaev-Zeldovich  (SZ) effect with the Planck satellite.  The SZ mass was estimated to be  $\sim 4  \times 10^{14} M_{\odot}$~\cite{Planck16}. Two diffuse radio sources have been identified at the cluster outskirts using  the LOFAR radio telescope~\cite{Rajpurohit25}. Such  symmetric radio relics are extremely rare and only six such systems have been observed in LOTSS-DR2~\cite{Jones23}. As part of a multi-wavelength observing campaign, deep X-ray~\cite{Stroe25}, radio~\cite{Rajpurohit25} and Weak lensing~\cite{Ahn25} observations  were obtained for this cluster using Subaru/Suprime-Cam (optical), uGMRT/VLA (radio), and  Chandra/XMM-Newton. The weak lensing analysis detected two distinct sub-clusters having masses ($M_{200}$) of $10^{14} M_{\odot}$ and $3 \times 10^{14} M_{\odot}$, with a separation of $\sim$~500 kpc~\cite{Ahn25}. To complement these multi-wavelength observations for this cluster, we do a targeted search for gamma-ray emission at GeV energies from this galaxy cluster using the latest data from the Fermi-LAT telescope.

This manuscript is organized as follows. Sect.~\ref{sec:da} describes the dataset and analysis methodology used for the search. The results found are shown in used is in Sect.~\ref{sec:results_1_300}. We conclude in Sect.~\ref{sec:conclusions}.

\section{Data Analysis}
\label{sec:da}
We searched for $\gamma$-ray emission from the galaxy cluster
PSZ2~G181.06+48.47~\cite{Planck2016} using observations from the Fermi Large Area Telescope (LAT), one of the two scientific
instruments onboard the Fermi Gamma-ray Space Telescope\footnote{\url{https://fermi.gsfc.nasa.gov/science/instruments/lat.html}}. The
LAT is a pair-conversion telescope with a field of view of
approximately $2.4~\mathrm{sr}$ and is sensitive to photons from
$\sim20~\mathrm{MeV}$ to beyond $300~\mathrm{GeV}$, making it well
suited for studies of high-energy emission from a wide variety of
astrophysical sources \citep{Atwood09,Fermilat2026}.
We analyzed nearly 18 years of Pass~8 SOURCE
(FRONT+BACK; \texttt{evclass=128}, \texttt{evtype=3}) data spanning the
Mission Elapsed Time (MET) interval
239587201--803865605, corresponding to observations obtained between
2008 August 5 and 2026 June 23. Events were
selected within a $5^\circ$ radius of the cluster centre
(RA=$144.85^\circ$, Dec=$40.75^\circ$) in the energy range
1--300~GeV. Energies below 1~GeV were excluded because the LAT
point-spread function becomes substantially broader at lower energies,
making the localization of cluster-scale emission more difficult.

The analysis was performed using the \texttt{Fermitools} package
(version 2.5.1) together with \texttt{Fermipy}
(version 1.4.2)~\cite{Wood2017} employing the \texttt{P8R3\_SOURCE\_V3} instrument response
functions. Standard data quality selections
(\texttt{DATA\_QUAL>0} and \texttt{LAT\_CONFIG==1}) were applied to
exclude periods of non-nominal spacecraft operation. In addition,
events with spacecraft rocking angles exceeding $52^\circ$ were
discarded, and a zenith-angle cut of $90^\circ$ was applied to suppress
contamination from Earth-limb $\gamma$ rays.
The data were spatially binned into pixels of
$0.2^\circ\times0.2^\circ$ and logarithmically divided into 10 energy
bins per decade over the 1--300~GeV energy range. The source model was
constructed from the FL16Y 16-year Fermi-LAT source catalogue~\cite{Fermilat2026}, together
with the standard Galactic diffuse emission model
\texttt{gll\_iem\_v07.fits} and the isotropic background model
\texttt{iso\_P8R3\_SOURCE\_V3\_v1.txt}\footnote{\url{https://fermi.gsfc.nasa.gov/ssc/data/access/lat/BackgroundModels.html}}. During the
likelihood optimization, the normalizations of both diffuse components
were allowed to vary freely, while the spectral parameters of catalogued
sources were treated according to the fitting strategy described in the
following section. Source detection and
spectral parameter estimation were performed using a standard binned
maximum-likelihood analysis based on the Poisson likelihood function.

\subsection{Binned Likelihood Analysis}

The $\gamma$-ray analysis was performed using the standard binned
maximum-likelihood method implemented in the Fermitools package. A
circular region of interest (ROI) of radius $5^\circ$ centred on
PSZ2~G181.06+48.47 was analysed throughout this work. As an initial
hypothesis, the cluster was modelled as a point source with a
PowerLaw spectrum and a photon index of $\Gamma=2.0$, which was
subsequently optimized during the likelihood fit.
The event selection was performed using the \texttt{gtselect} tool,
followed by the application of the standard good-time interval
selection with \texttt{gtmktime}. A two-dimensional counts map (CMAP)
was first generated using \texttt{gtbin} to verify the event
distribution within the ROI. The data were then binned into a
three-dimensional counts cube (CCUBE), which serves as the primary
input for the binned likelihood analysis.
The instrument exposure was computed using \texttt{gtltcube} and
\texttt{gtexpcube2}, producing the livetime cube and binned exposure
map required for the likelihood calculation. Expected counts from all
catalogued and diffuse sources were generated with
\texttt{gtsrcmaps}, after which the model parameters were optimized
using \texttt{gtlike}. During the likelihood optimization, the
normalizations of the Galactic and isotropic diffuse components and
the target source were allowed to vary, while catalogued background
sources were kept fixed. 

\subsection{Model Fitting using Maximum Likelihood}
\label{sec:mle}

The source parameters were determined using the standard maximum-likelihood
method implemented in the \texttt{gtlike} tool. For a given source model,
\texttt{gtlike} maximizes the likelihood of observing the measured
$\gamma$-ray data by optimizing the free spectral parameters while
accounting for the instrumental response, diffuse backgrounds, and all
catalogued sources included in the model.
To search for previously unmodelled emission and to evaluate its
statistical significance, residual Test Statistic (TS) maps were
generated using the \texttt{gttsmap} tool. At each pixel of the map, a
test source is introduced and its normalization is optimized while the
likelihood is recomputed. The TS is defined as

\begin{equation}
TS = 2\left(\log\mathcal{L}_{1}-\log\mathcal{L}_{0}\right),
\end{equation}

where $\mathcal{L}_{0}$ and $\mathcal{L}_{1}$ are the maximum
likelihoods obtained for the null hypothesis (without the test source)
and the alternative hypothesis (including the test source),
respectively. Under the conditions of Wilks' theorem, the TS follows
approximately a half-$\chi^{2}$ distribution for a source whose flux is
constrained to be non-negative \citep{Mattox96}. Consequently, the
detection significance is approximately given by
$\sqrt{TS}$ for sufficiently large photon statistics.

\section{Results in the 1--300~GeV Energy Range}
\label{sec:results_1_300}

We first analyzed the region surrounding PSZ2~G181.06+48.47 using the
baseline FL16Y source model without including any additional source at
the cluster position. This reference model provides the baseline against
which all subsequent spatial and spectral analyses are compared.

\subsection{Point-Source Spatial Model}
The baseline model comprised all FL16Y 16-year catalog sources
within the ROI~\citep{Fermilat2026}, together with the Galactic diffuse
emission model (\texttt{gll\_iem\_v07}) and the isotropic diffuse
background (\texttt{iso\_P8R3\_SOURCE\_V3\_v1}). The corresponding
residual TS map is shown in Fig.~\ref{fig:tsmap_comparison}(a). A
significant excess is detected at the cluster position with
$\mathrm{TS}=25.9$ ($\approx5.1\sigma$), indicating that the baseline
catalog model does not fully account for the observed emission.
A second, brighter excess is also present approximately
$1.58^{\circ}$ north of the cluster, with
$\mathrm{TS}_{\max}=45.3$ ($\approx6.7\sigma$) at
$(\alpha,\delta)=(145.53^{\circ},\,42.25^{\circ})$.
The detection of these residual excesses motivated the introduction of
an additional source at the cluster position, initially modelled as a
point source.

\subsubsection{Addition of the Target Source}
An additional point source was introduced at the cluster position,
$(\alpha,\delta)=(144.85^{\circ},\,40.75^{\circ})$, with an initial
PowerLaw spectrum ($\Gamma=2.0$), and the model was refitted using the
\texttt{NEWMINUIT} optimizer\footnote{See \url{https://fermi.gsfc.nasa.gov/ssc/data/analysis/likelihood/fitting_models.html}}. The likelihood fit yielded a best-fitting
photon index of $\Gamma=3.20\pm0.62$, an integrated photon flux of
$(9.75\pm2.74)\times10^{-11}\,\mathrm{ph\,cm^{-2}\,s^{-1}}$ over the
1--300~GeV energy range, and a source detection significance of
$\mathrm{TS}=19.0$, corresponding to approximately
$4.4\sigma$.

\subsubsection{Residual Emission}
\begin{figure*}[htbp]
    \centering
    \begin{minipage}{0.49\textwidth}
        \centering
        \includegraphics[width=\linewidth]{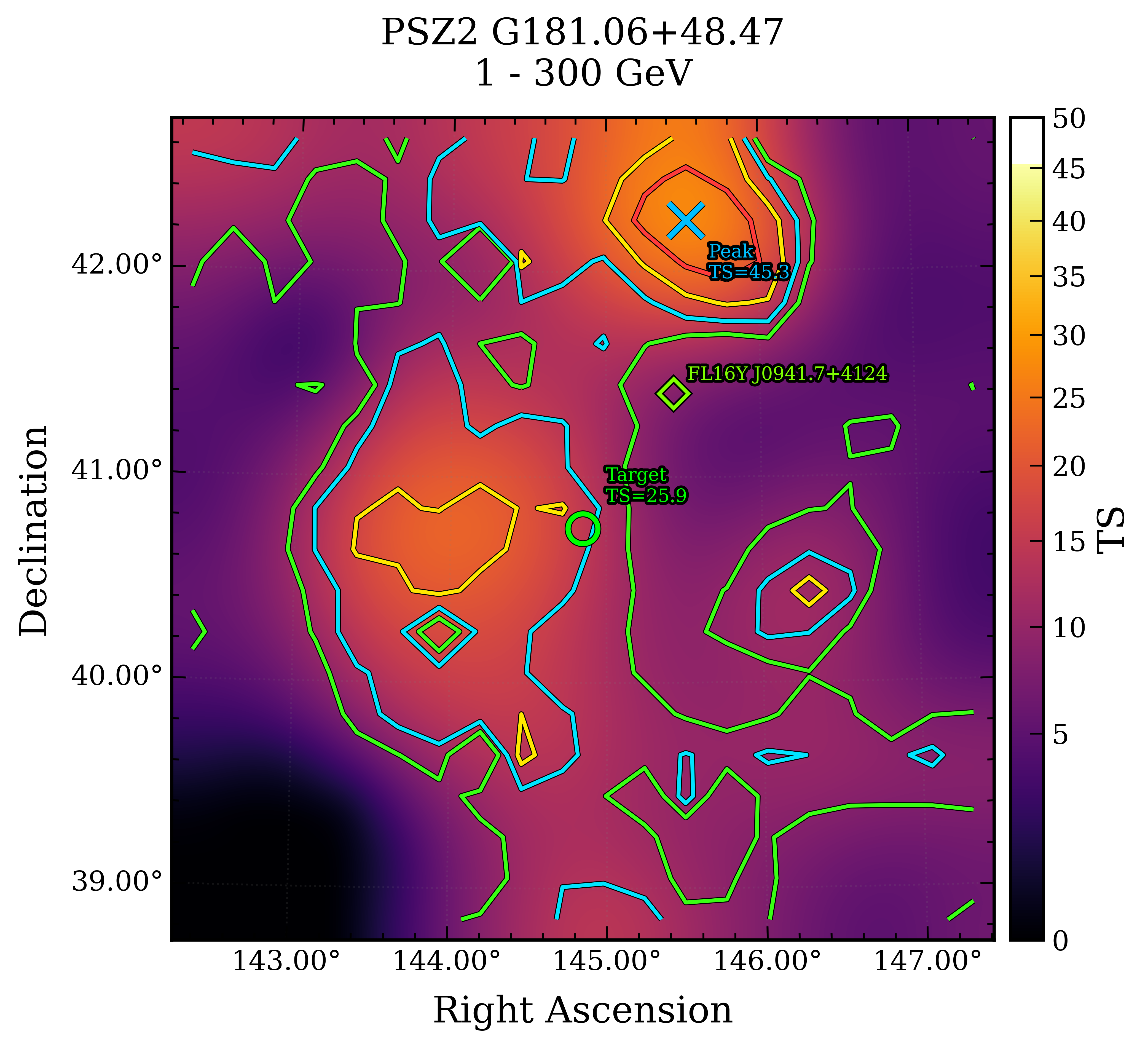}\\[2pt]
        \textbf{(a)} Baseline model (no source at target position).
    \end{minipage}
    \hfill
    \begin{minipage}{0.49\textwidth}
        \centering
        \includegraphics[width=\linewidth]{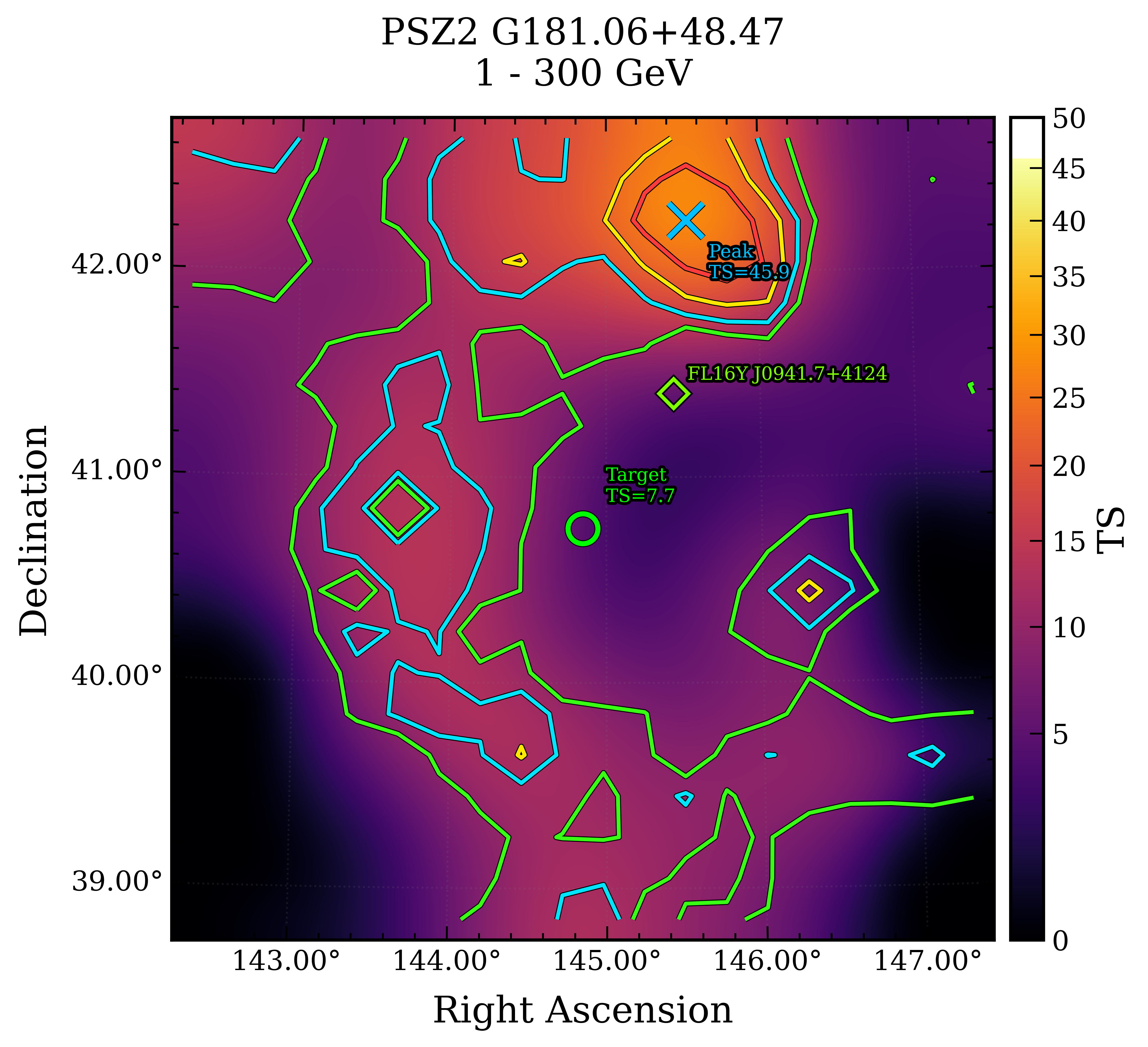}\\[2pt]
        \textbf{(b)} Residual TS map after adding the target source.
    \end{minipage}
    \caption{Comparison of TS maps of the region surrounding
    PSZ2~G181.06+48.47 in the 1--300~GeV energy band. The green circle
    marks the target cluster position, and the blue cross marks the
    pixel of maximum TS. Contours are drawn on the unsmoothed TS map at
    TS $= 9, 16, 25, 36$, shown in green, cyan, yellow, and red,
    respectively. The nearby green diamond indicates the nearest catalog source
    from the ROI background model. The field of view shown is
    $\sim4^\circ \times 4^\circ$ in both panels.
    (\textbf{a}) TS map obtained using the baseline model consisting of
    the FL16Y catalog and diffuse backgrounds only. A significant
    excess is observed at the cluster position with $\mathrm{TS}=25.9$,
    while the brightest excess in the ROI has $\mathrm{TS}=45.3$ at
    $(145.53^{\circ},\,42.25^{\circ})$.
    (\textbf{b}) Residual TS map after adding a point source at the
    cluster position and performing a maximum-likelihood fit. The
    residual TS at the cluster decreases to $7.7$, indicating that the
    added source accounts for most of the emission associated with the
    cluster, whereas the nearby excess remains the dominant residual
    ($\mathrm{TS}=45.9$).}
    \label{fig:tsmap_comparison}
\end{figure*}

A residual TS map was generated using the best-fitting point-source
model (Fig.~\ref{fig:tsmap_comparison}b). Following the inclusion of the
target source, the residual TS at the cluster position decreases from
25.9 to 7.7, indicating that the point-source model accounts for a
substantial fraction of the emission associated with
PSZ2~G181.06+48.47. Nevertheless, significant residual emission remains
at the cluster position, suggesting that the point-source hypothesis
does not fully describe the observed $\gamma$-ray emission. In
addition, the bright offset excess persists essentially unchanged, with
$\mathrm{TS}_{\max}=45.9$ at
$(\alpha,\delta)=(145.53^{\circ},\,42.25^{\circ})$. The persistence of
this hotspot motivates a more detailed investigation of the spatial
morphology of the cluster emission and the nature of the neighbouring
excess in the following sections.

\subsection{Spatial Extension Analysis}
\label{sec:extension}
Galaxy clusters are intrinsically extended systems and therefore may not
be adequately described by a point-source morphology. Motivated by the
residual emission remaining at the cluster position after fitting the
point-source model, we investigated whether an extended spatial template
provides a statistically better description of the observed $\gamma$-ray emission\footnote{\url{https://fermi.gsfc.nasa.gov/ssc/data/analysis/scitools/xml_model_defs.html}}.

\subsubsection{Radial Gaussian Spatial Model}

The point-source spatial model was replaced by a
\texttt{RadialGaussian} template while keeping the centroid fixed at the
cluster position,
$(\alpha,\delta)=(144.85^{\circ},\,40.75^{\circ})$.
The Gaussian width was fixed during each fit and values of
$\sigma=0.1^{\circ}$,
$0.2^{\circ}$,
$0.3^{\circ}$,
$0.4^{\circ}$ and
$0.5^{\circ}$ were investigated.
For each trial, new source maps were generated and the complete
likelihood analysis was repeated using the same analysis configuration
adopted for the point-source model.

The results obtained for the Gaussian spatial templates are summarized
in Table~\ref{tab:gaussian_extension}.
The corresponding source detection significances were computed relative
to the null hypothesis without the target source, while the extension
test statistic was evaluated as
\begin{equation}
TS_{\rm ext}
=
2\left(
\log\mathcal{L}_{\rm Gaussian}
-
\log\mathcal{L}_{\rm Point}
\right).
\end{equation}

The quantity $TS_{\rm ext}$ is a likelihood-ratio statistic
used for model selection between competing spatial hypotheses in
Fermi-LAT analyses. It measures the improvement in the maximum
likelihood obtained by replacing the point-source model with an extended
Gaussian template. In the present analysis, both spatial models have the
same number of free parameters, as only the spectral normalization and
photon index are optimized during each likelihood fit, while the source
position and Gaussian width are kept fixed. Consequently, the comparison
is not affected by differences in model complexity, and
$TS_{\rm ext}$ provides a direct and unbiased measure of the statistical
preference for the Gaussian spatial model over the point-source
hypothesis.

\begin{table*}[htbp]
\centering
\caption{Likelihood results obtained using RadialGaussian spatial
templates. The extension test statistic,
$TS_{\rm ext}$, is computed directly from the fitted likelihood values
as
$TS_{\rm ext}=2(\log\mathcal{L}_{\rm Gaussian}-\log\mathcal{L}_{\rm Point})$.
Consequently, $TS_{\rm ext}$ is not expected to equal the difference
between the reported source detection significances
($TS_{\rm source}$), since the latter are evaluated relative to the
null hypothesis without the target source and are reported separately
by the likelihood optimizer.}
\label{tab:gaussian_extension}
\begin{tabular}{cccccc}
\hline
Model &
$-\log\mathcal{L}$ &
$TS_{\rm source}$ &
$TS_{\rm ext}$ &
Photon Index &
Photon Flux ($10^{-10}$ ph cm$^{-2}$ s$^{-1}$)\\
\hline
Point Source &
49414.913 &
19.1 &
0 &
$3.20\pm0.62$ &
$0.975\pm0.274$\\

$\sigma=0.1^\circ$ &
49412.251 &
23.6 &
5.3 &
$2.89\pm0.43$ &
$1.165\pm0.293$\\

$\sigma=0.2^\circ$ &
49408.904 &
30.3 &
12.0 &
$2.80\pm0.38$ &
$1.608\pm0.343$\\

$\sigma=0.3^\circ$ &
49403.781 &
40.5 &
22.3 &
$2.68\pm0.32$ &
$2.281\pm0.413$\\

$\sigma=0.4^\circ$ &
49396.525 &
55.0 &
36.8 &
$2.62\pm0.26$ &
$3.220\pm0.496$\\

$\sigma=0.5^\circ$ &
49425.280 &
-- &
-- &
$2.71\pm0.08$ &
$4.699\pm0.346$\\
\hline
\end{tabular}
\end{table*}

A monotonic improvement in the fit quality is observed as the Gaussian
width increases from $\sigma=0.1^{\circ}$ to
$\sigma=0.4^{\circ}$.
The corresponding source detection significance increases from
$TS_{\rm source}=23.6$ to
$TS_{\rm source}=55.0$ ($\approx7.4\sigma$),
while the extension statistic reaches
$TS_{\rm ext}=36.8$.
Simultaneously, the best-fitting spectrum becomes progressively harder,
with the photon index changing from
$\Gamma=3.20\pm0.62$
for the point-source model to
$\Gamma=2.62\pm0.26$
for the
$\sigma=0.4^{\circ}$ Gaussian model.
The corresponding integrated photon flux increases from
$9.75\times10^{-11}$
to
$3.22\times10^{-10}\,
\mathrm{ph\,cm^{-2}\,s^{-1}}$. For $\sigma=0.5^{\circ}$, the likelihood optimization fails to converge
to a physically meaningful solution. The Galactic diffuse normalization decreases to
$5.66\times10^{-7}$,
its spectral index becomes effectively unconstrained
($0.59\pm87.36$),
while the isotropic diffuse normalization increases to 1.32.
These values are physically implausible, indicating that the optimizer
has converged to an unphysical local minimum.
Consequently, the
$\sigma=0.5^{\circ}$ model is rejected and is not considered further.

\subsubsection{Radial Disk Spatial Model}
\label{sec:disk_extension}
To investigate whether the observed emission is better described by a
uniformly extended morphology, the point-source model was replaced by a
\texttt{RadialDisk} spatial template while keeping the centroid fixed at
the cluster position,
$(\alpha,\delta)=(144.85^{\circ},\,40.75^{\circ})$.
The disk radius was fixed during each likelihood fit and varied over the
range $R=0.1^{\circ}$--$2.0^{\circ}$. For each trial radius, new source
maps were generated using \texttt{gtsrcmaps}, followed by a complete
binned likelihood fit using \texttt{gtlike}. The preferred extension was
therefore determined from the likelihood profile. The extension test statistic was computed relative to the point-source
model as
\begin{equation}
TS_{\rm ext}
=
2\left(
\log\mathcal{L}_{\rm Disk}
-
\log\mathcal{L}_{\rm Point}
\right).
\end{equation}

The results are summarized in
Table~\ref{tab:disk_extension}.

\begin{table*}[htbp]
\centering
\caption{Likelihood profile obtained using RadialDisk spatial templates.}
\label{tab:disk_extension}
\begin{tabular}{cccccc}
\hline
Radius
(deg) &
$-\log\mathcal{L}$ &
$TS_{\rm source}$ &
$TS_{\rm ext}$ &
Photon Index &
Photon Flux
($10^{-10}$ ph cm$^{-2}$ s$^{-1}$)\\
\hline

Point &
49414.9 &
19.0 &
0 &
$3.20\pm0.62$ &
$0.98\pm0.27$\\

0.1 &
49413.9 &
20.2 &
1.9 &
$3.01\pm0.49$ &
$1.03\pm0.28$\\

0.2 &
49411.9 &
24.3 &
6.0 &
$2.85\pm0.39$ &
$1.17\pm0.29$\\

0.3 &
49410.7 &
26.6 &
8.4 &
$2.90\pm0.42$ &
$1.37\pm0.32$\\

0.4 &
49409.4 &
29.3 &
11.1 &
$2.88\pm0.41$ &
$1.61\pm0.34$\\

0.5 &
49407.0 &
34.1 &
15.8 &
$2.78\pm0.38$ &
$1.94\pm0.38$\\

0.8 &
49397.1 &
54.1 &
35.8 &
$2.67\pm0.27$ &
$3.30\pm0.50$\\

1.0 &
49386.4 &
75.2 &
57.0 &
$2.63\pm0.22$ &
$4.63\pm0.59$\\

1.1 &
49379.3 &
89.4 &
71.2 &
$2.60\pm0.19$ &
$5.43\pm0.64$\\

1.5 &
49349.6 &
148.8 &
130.6 &
$2.60\pm0.15$ &
$8.96\pm0.80$\\

2.0 &
49316.1 &
216.0 &
197.8 &
$2.62\pm0.12$ &
$14.33\pm1.05$\\

\hline
\end{tabular}
\end{table*}

All RadialDisk fits converged successfully, with physically reasonable
Galactic and isotropic diffuse background normalizations throughout the
scan, indicating stable likelihood optimization. The likelihood improves monotonically as the disk radius increases,
without reaching a well-defined maximum.
Consequently, both
$TS_{\rm source}$
and
$TS_{\rm ext}$
increase continuously with increasing disk radius.
The best-fitting photon index gradually hardens, while the inferred
photon flux increases by more than an order of magnitude.

Residual TS maps were generated for the largest disk radii to determine
whether the increasing likelihood resulted from a better description of
the cluster emission or from the absorption of neighbouring structures.
For
$R=1.0^{\circ}$,
$1.1^{\circ}$,
and
$1.5^{\circ}$,
the residual TS at the cluster position remained low
($TS_{\rm res}\lesssim2$),
while the neighbouring hotspot remained clearly detectable with peak
residual TS values of approximately
46,
44,
and
31,
respectively. For
$R=2.0^{\circ}$,
however,
the neighbouring hotspot is largely absorbed,
its peak residual TS decreasing to only
20.5,
while the brightest residual shifts to a different location. 

Consequently, although the RadialDisk model formally provides a higher
likelihood than the point-source model, the absence of a well-defined
maximum in the likelihood profile together with the progressive
suppression of neighbouring emission indicates that it does not recover
a physically meaningful extent for
PSZ2~G181.06+48.47. We therefore do not adopt the RadialDisk model in
the subsequent analysis and instead use the
$\sigma=0.4^{\circ}$ RadialGaussian model as the preferred spatial
representation of the cluster emission.

\subsubsection{Residual TS Map for the Preferred Spatial Model}
\label{sec:residual_extended}
To verify that the improvement obtained using the preferred extended
spatial model is genuinely associated with
PSZ2~G181.06+48.47 rather than resulting from neighboring sources,
a residual TS map was generated using \texttt{gttsmap} after refitting
the best-fitting RadialGaussian model
($\sigma=0.4^{\circ}$)
with \texttt{gtlike}. The resulting TS map represents the significance
of any remaining excess after subtraction of the complete best-fit model. The best-fitting RadialGaussian template yields
$TS_{\rm source}=55.0$,
corresponding to a substantial improvement over the point-source
hypothesis. The quantitative properties of the residual emission are summarized in
Table~\ref{tab:residual_comparison}.

\begin{table}[htbp]
\centering
\caption{Comparison of residual TS values obtained using different
models for PSZ2~G181.06+48.47.}
\label{tab:residual_comparison}
\begin{tabular}{lcc}
\hline
Model &
Residual TS at Target &
Peak Residual TS \\
\hline
No Target Source & 25.9 & 45.3 \\
Point Source & 7.7 & 45.9 \\
RadialGaussian ($\sigma=0.4^\circ$) & $\approx0.00$ & 46.2 \\
\hline
\end{tabular}
\end{table}

The residual TS at the cluster position decreases dramatically as
progressively more realistic source models are adopted.
Without including the cluster in the model, the residual TS at the
target position is 25.9.
Introducing a point source reduces this value to 7.7, demonstrating
that the point-source hypothesis accounts for a significant fraction of
the observed emission but leaves a statistically significant residual at
the cluster center. Replacing the point source with the best-fitting RadialGaussian template
almost completely removes the central residual, yielding a residual TS
consistent with zero. This demonstrates that the observed emission is
better described by an extended Gaussian morphology than by a
point-like source. Importantly, the brightest residual excess in the ROI remains
essentially unchanged after introducing the extended source model.
The peak residual TS is
46.2,
located at
$(\alpha,\delta)\approx(145.53^{\circ},\,42.25^{\circ})$,
approximately
$1.58^{\circ}$
from the cluster center.
Both the location and significance of this excess remain unchanged
within the pixel resolution of the TS map, demonstrating that the
Gaussian template does not absorb the neighboring residual source. These results demonstrate that the improvement obtained with the
RadialGaussian model arises from a better description of the emission
associated with PSZ2~G181.06+48.47 rather than from artificially
absorbing neighbouring residual emission. The persistence of the offset
hotspot therefore supports its interpretation as an independent
$\gamma$-ray source, while the cluster emission itself is more
accurately represented by an extended Gaussian morphology than by a
point source.

\begin{figure}[htbp]
\centering
\includegraphics[width=0.85\linewidth]{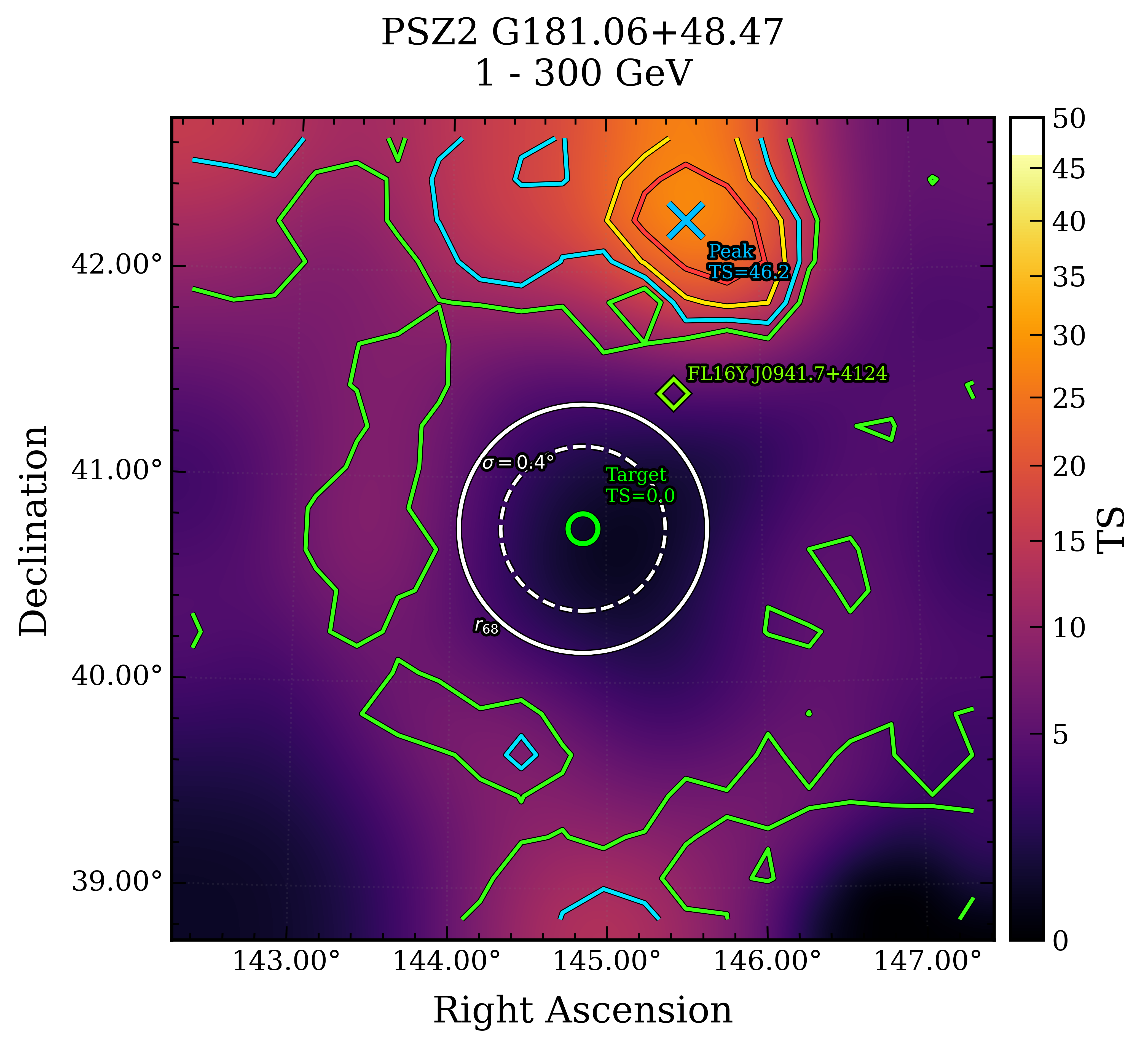}
\caption{
Residual TS map obtained after fitting
PSZ2~G181.06+48.47
with the preferred RadialGaussian spatial model
($\sigma=0.4^\circ$).
The white dashed circle marks the best-fit Gaussian's standard
deviation ($\sigma=0.4^\circ$), and the white solid circle marks the
corresponding 68\% containment radius
($r_{68}=1.51\sigma\simeq0.604^\circ$).
Green, cyan, yellow, and red contours correspond to
TS~$=9,16,25,36$, respectively.
The residual emission at the cluster position is almost completely
removed, whereas the prominent hotspot located approximately
$1.58^\circ$
from the cluster center remains essentially unchanged.}
\label{fig:gaussian_residual}
\end{figure}

\subsubsection{Simultaneous Modeling of the Cluster and the Nearby Gamma-Ray Hotspot}
\label{sec:two_source_model}
The TS maps presented in the previous section demonstrated that the
$\gamma$-ray emission associated with PSZ2~G181.06+48.47 is well
described by a RadialGaussian template with
$\sigma=0.4^{\circ}$, reducing the TS at the cluster position to
a value consistent with zero. However, a prominent nearby hotspot with
$\mathrm{TS}\approx46$ remained at
$(\alpha,\delta)=(145.53^{\circ},\,42.25^{\circ})$,
approximately $1.58^{\circ}$ north of the cluster center. This position lies $0.31^{\circ}$ from
4FGL~J0943.6+4207 ($\alpha=145.910^{\circ}$, $\delta=+42.117^{\circ}$),
the nearest entry in the 4FGL-DR4 catalog~\cite{4FGL-DR4}, which is listed there with
$\mathrm{TS}=18.8$ and a PowerLaw spectrum. 
To identify a plausible multiwavelength counterpart, we queried SIMBAD
for objects in the vicinity of 4FGL~J0943.6+4207. The catalog itself
lists the source as a $\gamma$-ray source (\texttt{gam}, 4FGL) with a
candidate-quasar counterpart, while the nearest object overall in the
SIMBAD search is the unresolved radio source
NVSS~J094204+421422, located only $36.3^{\prime\prime}$ from the 4FGL
position---essentially coincident with it given the LAT point-spread
function at these energies~\cite{Shimwell2022}.
We further note that the TS recovered in our re-analysis ($\approx 46$) is substantially higher than the value quoted in the
4FGL catalog (18.8), consistent with the longer effective exposure of
the 16-year data set used here relative to the 4FGL-DR4 integration
period. Such a compact radio source at the nominal position of an
otherwise unassociated \emph{Fermi-LAT} source is the signature commonly
seen for blazar-class active galactic nuclei, which dominate the
identified extragalactic 4FGL population. Also present within the search radius, though at larger offsets ($5^\prime$--$14^\prime$), are several Seyfert-type AGN (Sy2: SDSS~J094131.19+421214.8; Sy1: SDSS~J094058.75+420855.4) and, notably, a radio galaxy, FIRST~J094101.3+421511, at $11.9^\prime$. Radio galaxies and Seyferts are
established, if less common, gamma-ray-emitting AGN classes, and their
presence in the field means a definitive counterpart identification for
4FGL~J0943.6+4207 cannot be made from position alone; we flag
NVSS~J094204+421422 as the most positionally compelling candidate but
regard the counterpart question as open pending dedicated
multiwavelength follow-up. 

We further cross-matched this hotspot position against every source in
the FL16Y 16-year source list used to construct the initial background
model, by angular separation. The nearest FL16Y entry is
FL16Y~J0941.7+4124 ($\mathrm{TS}=118.1$), $0.84^{\circ}$ away -- i.e.
4FGL~J0943.6+4207 itself has no counterpart in the FL16Y catalog,
despite being a firmly-detected 4FGL source at this position. A likely
explanation is that its comparatively low 4FGL TS (18.8) placed it below
the significance threshold used to construct the FL16Y source list, out
of the $\sim7201$ sources it contains. Since our re-analysis recovers a
substantially higher significance at this position ($\mathrm{TS}\approx46$
in the single-source residual map, and formally detected at high
significance in the simultaneous two-source fit below), this source
should be added to the FL16Y model at its correct 4FGL coordinates
($\alpha=145.910^{\circ}$, $\delta=+42.117^{\circ}$) as a new source
rather than treated as a residual artifact of the cluster fit. Its
absence from the FL16Y background model explains why it appears as an
unmodeled residual hotspot in the maps of the previous section. To investigate whether this neighboring hotspot influences the inferred
cluster emission, an additional RadialGaussian source was introduced at
its position while retaining the
$\sigma=0.4^{\circ}$ Gaussian model for the cluster. Both sources were
described by power-law spectra and were fitted simultaneously together
with the Galactic and isotropic diffuse backgrounds.
The simultaneous fit converged successfully with a significantly improved
likelihood, $-\log\mathcal{L}=49356.4$,
representing a substantial improvement over the single-Gaussian model
($-\log\mathcal{L}=49396.5$). 

\begin{table}[htbp]
\centering
\caption{Best-fitting spectral parameters obtained from the simultaneous likelihood fit in which both PSZ2~G181.06+48.47 and the neighbouring $\gamma$-ray hotspot are modelled using RadialGaussian spatial templates with fixed width $\sigma=0.4^{\circ}$. The table lists the photon index and integrated photon flux ($1$--$300$~GeV) derived for each source.}
\label{tab:dual_gaussian_spectra}
\begin{tabular}{lcc}
\hline
Source &
Photon Index &
Photon Flux
($10^{-10}$ ph cm$^{-2}$ s$^{-1}$) \\
\hline
Cluster &
$2.60\pm0.26$ &
$3.15\pm0.49$ \\

Nearby hotspot &
$2.66\pm0.22$ &
$3.71\pm0.48$ \\
\hline
\end{tabular}
\end{table}

\begin{figure*}[htbp]
\centering
\includegraphics[width=0.72\textwidth]{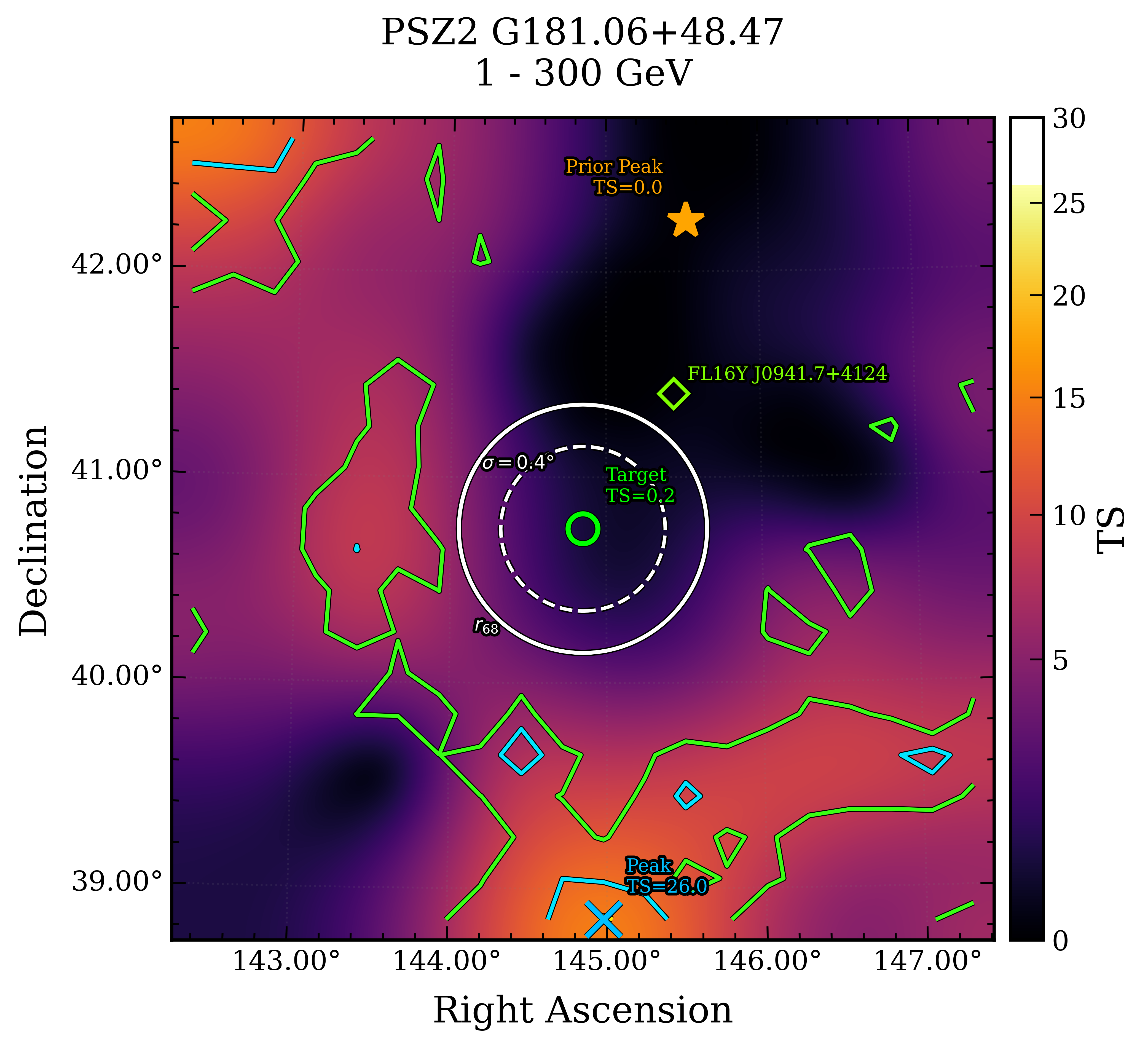}
\caption{
Residual TS map obtained after simultaneously fitting the
$\sigma=0.4^{\circ}$ RadialGaussian model for
PSZ2~G181.06+48.47 together with an additional RadialGaussian source at
the position of the neighboring residual hotspot. The green circle marks
the cluster position, while the orange diamond marks the centroid of the
second source. The residual TS at both source positions is consistent
with zero ($TS_{\rm cluster}=0.2$ and
$TS_{\rm residual}\approx0$), demonstrating that the two-source model
successfully accounts for the dominant $\gamma$-ray emission in the ROI.
The brightest remaining residual ($TS_{\max}=26.0$) is located nearly
$2^{\circ}$ south of the cluster and is therefore unlikely to affect the
derived properties of PSZ2~G181.06+48.47.
}
\label{fig:two_source_residual}
\end{figure*}

A residual TS map was subsequently generated using the best-fit
two-source model (Fig.~\ref{fig:two_source_residual}). The residual TS at
the cluster position is $TS_{\rm cluster}=0.2$,
while the residual at the position of the neighboring hotspot is
$TS_{\rm hotspot}\approx 0$.
Thus, both principal excesses identified in the ROI are completely
accounted for by the two-source model. The brightest remaining excess in
the residual map is located at
$(\alpha,\delta)=(144.98^{\circ},\,38.85^{\circ})$,
with
$TS_{\max}=26.0$,
approximately $1.9^{\circ}$ south of the cluster center. The nearest
FL16Y catalog entry to this position is FL16Y~J0934.2+3927
($\mathrm{TS}=272.9$), $1.25^{\circ}$ away, confirming that this excess
also has no counterpart in the background model. As it is spatially well
separated from both the cluster and the neighboring hotspot, it is
unlikely to influence the derived properties of PSZ2~G181.06+48.47.

For completeness, we also cross-matched the cluster center itself,
$(\alpha,\delta)=(144.85^{\circ},\,40.75^{\circ})$, against the FL16Y
catalog; the nearest entry, FL16Y~J0941.7+4124, lies $0.73^{\circ}$
away, ruling out a chance coincidence with an unmodeled point source at
the cluster position.

\paragraph{Morphology of the neighboring hotspot.}
Having established that the two-source model removes both excesses, we
next tested whether the neighboring hotspot is itself better described
by an extended template or by a point source, and whether this choice
biases the recovered properties of the cluster. Two variants of the
simultaneous fit were run, identical in every respect except for the
spatial model assigned to the neighboring source: (i) a RadialGaussian
with $\sigma=0.4^{\circ}$ (as adopted above), and (ii) a point source
(\texttt{SkyDirFunction}) at the same position. The
point-source variant converged to $-\log\mathcal{L}=49373.4$, compared with $-\log\mathcal{L}=49356.43$ for the Gaussian variant, a
difference of
$\Delta(-\log\mathcal{L}) = 49373.4-49356.4 = 17.0$,
corresponding to
$TS_{\rm ext} = 2\,\Delta(-\log\mathcal{L}) \approx 33.9$,
in favor of the extended (Gaussian) description of the neighboring
source. The residual TS maps generated from each variant reinforce this:
the point-source fit leaves a residual of $TS_{\rm residual}=2.8$ at
the source position, whereas the Gaussian fit leaves
$TS_{\rm residual}\approx0$, indicating that the point source is unable
to fully absorb the excess and that the emission is intrinsically more
extended than the LAT point-spread function at these energies. The two
variants also differ markedly in their recovered spectra: the
point-source fit is both softer and fainter than the Gaussian fit,a pattern consistent with the point source compensating for flux spread
over a region larger than the PSF by steepening the spectrum and
lowering the normalization, rather than fully capturing the excess.

\begin{table}[htbp]
\centering
\caption{Best-fitting spectral parameters of the neighbouring $\gamma$-ray hotspot for different spatial models. The Gaussian model ($\sigma=0.4^{\circ}$) provides a significantly better description of the neighbouring source than a point-source hypothesis.}
\label{tab:neighbor_model}
\begin{tabular}{lcc}
\hline
Neighboring-source model &
Photon Index &
Photon Flux
($10^{-10}$ ph cm$^{-2}$ s$^{-1}$) \\
\hline
Point source &
$3.27\pm0.45$ &
$1.60\pm0.29$ \\

Gaussian ($\sigma=0.4^{\circ}$) &
$2.66\pm0.22$ &
$3.71\pm0.48$ \\
\hline
\end{tabular}
\end{table}

Crucially, the cluster's best-fit spectral parameters are consistent
within $1\sigma$ between the two variants, $(2.60\pm0.26)$ and
$(3.15\pm0.49)\times10^{-10}\,{\rm ph\,cm^{-2}\,s^{-1}}$ for the Gaussian
neighbouring-source variant versus $2.59$ and
$3.22\times10^{-10}\,{\rm ph\,cm^{-2}\,s^{-1}}$ for the point-source
variant, agreeing within their uncertainties. The assumed morphology of
the neighboring source therefore does not significantly bias the
inferred properties of the cluster. We caution against interpreting this result as an extension measurement
of 4FGL~J0943.6+4207 itself: the Gaussian template was introduced after
inspection of the residual map, making this an exploratory morphological
comparison rather than a formal extension analysis of that source. What
the comparison does robustly support is the more limited statement that,
within our ROI and background model, the residual excess coincident with
4FGL~J0943.6+4207 is more accurately described by a Gaussian spatial
template than by a point source, improving the likelihood by
$\Delta TS_{\rm ext}\approx34$, while leaving the cluster's derived
properties unaffected. These results demonstrate that the extended emission associated with the
cluster is not produced by contamination from the neighboring
$\gamma$-ray source, and that this conclusion is robust to the assumed
spatial model of that neighboring source. Instead, the data are best
described by two independent emitters: an extended source spatially
coincident with PSZ2~G181.06+48.47 and a second, previously
FL16Y-unmodeled $\gamma$-ray source, identified with
4FGL~J0943.6+4207, located approximately $1.58^{\circ}$ to the north
and itself preferring an extended rather than point-like description.

\subsubsection{Spectral Energy Distribution}
\label{sec:sed}

To characterize the broadband spectral behavior of the target, we
computed the spectral energy distribution (SED) using
\texttt{easyFermi}~\citep{2022A&C....4000609D}, dividing the
1--300~GeV range into five logarithmically spaced energy bins. In each
bin, an independent likelihood fit was performed with the normalization
of the target source left free, while the spectral shape and
normalization of all other sources in the ROI were fixed to their
best-fit values from the global analysis. Bins with a detection
significance below $3\sigma$ (TS~$<9$) are reported as $95\%$
confidence-level upper limits.

\begin{figure}[htbp]
    \centering
    \includegraphics[width=0.9\textwidth]{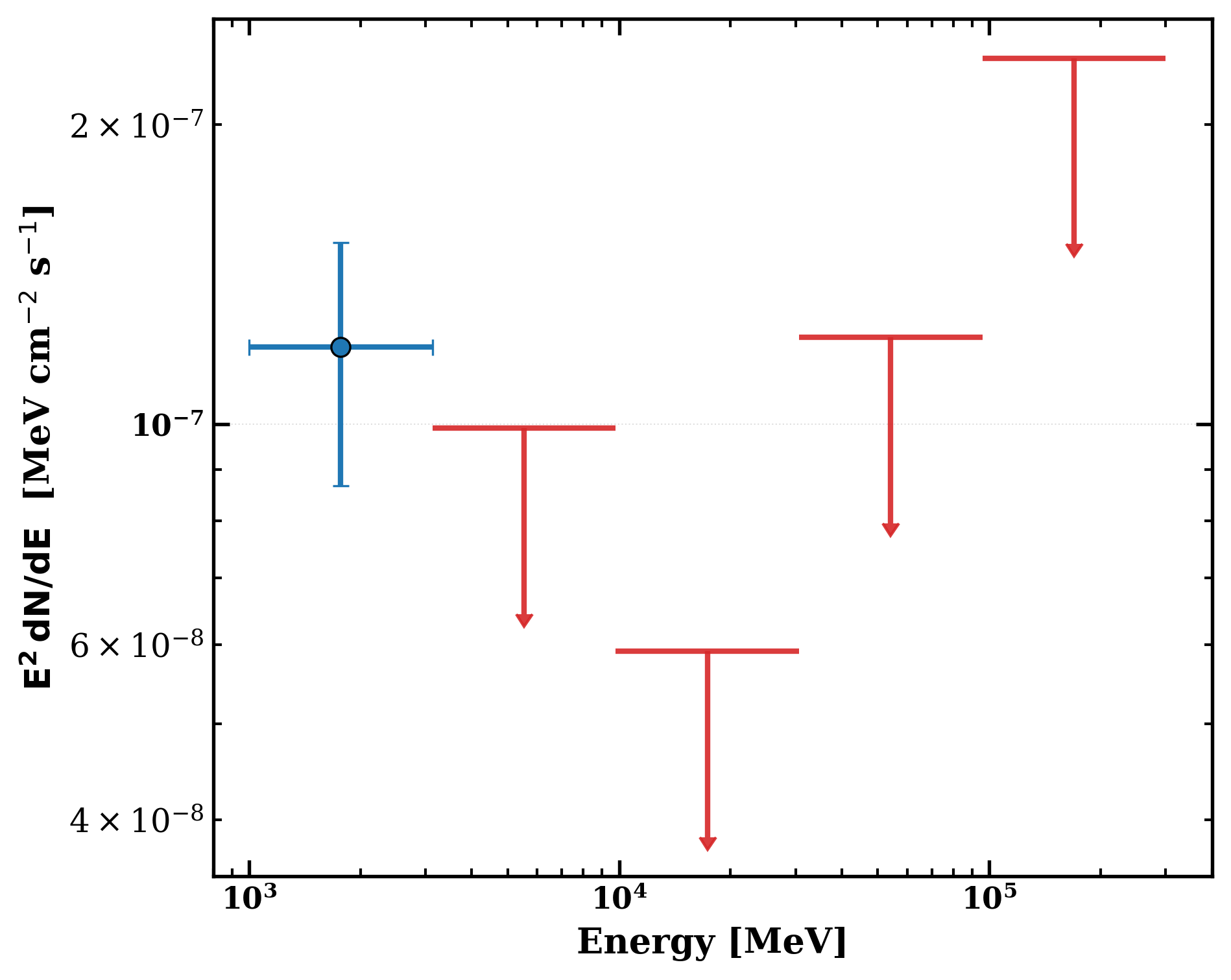}
    \caption{Spectral energy distribution (SED) of PSZ2~G181.06+48.47
    obtained with \texttt{easyFermi} over the 1--300~GeV range, using
    five logarithmically spaced energy bins. The blue filled circle
    marks the only bin with a significant detection
    (TS~$\geq9$, $\approx3\sigma$), with vertical and horizontal error
    bars showing the $1\sigma$ statistical uncertainty and the bin
    width, respectively. Red downward arrows denote $95\%$
    confidence-level upper limits for the remaining bins (TS~$<9$),
    with horizontal bars again indicating the bin width.}
    \label{fig:sed}
\end{figure}

The resulting SED is shown in Fig.~\ref{fig:sed}, with the numerical
values summarized in Table~\ref{tab:sed_numerical_values}. Only the
lowest energy bin, spanning $1.0$--$3.13$~GeV, yields a significant
detection, with $\mathrm{TS}=19.0$ ($\approx4.4\sigma$) and a flux of
$E^2\,dN/dE = (1.20\pm0.33)\times10^{-7}~\mathrm{MeV\,cm^{-2}\,s^{-1}}$. The second bin ($3.13$--$9.79$~GeV)
falls below the detection threshold, with $\mathrm{TS}=4.8$
($\approx2.2\sigma$), yielding a $95\%$ upper limit of
$9.9\times10^{-8}~\mathrm{MeV\,cm^{-2}\,s^{-1}}$. The remaining three
bins, covering $9.79$--$300$~GeV, show no evidence of emission
($\mathrm{TS}=0.00$ in all three bins), with corresponding upper limits
of $5.9\times10^{-8}$, $1.2\times10^{-7}$, and
$2.3\times10^{-7}~\mathrm{MeV\,cm^{-2}\,s^{-1}}$ respectively.

\begin{table}[htbp]
\centering
\caption{Bin boundaries, flux measurements, and associated TS values
for PSZ2~G181.06+48.47.}
\label{tab:sed_numerical_values}
\begin{tabular}{ccccc}
\toprule
Bin & $E_{\min}$ & $E_{\max}$ &
$E^2\,dN/dE$ Flux / 95\% U.L. & TS \\
 & (MeV) & (MeV) &
($\mathrm{MeV\,cm^{-2}\,s^{-1}}$) & Value \\
\midrule
1 & 1{,}000  & 3{,}129   & $(1.20\pm0.33)\times10^{-7}$ & 19.0 \\
2 & 3{,}129  & 9{,}791   & $<9.9\times10^{-8}$          & 4.8 \\
3 & 9{,}791  & 30{,}638  & $<5.9\times10^{-8}$          & 0.0  \\
4 & 30{,}638 & 95{,}873  & $<1.2\times10^{-7}$          & 0.0  \\
5 & 95{,}873 & 300{,}000 & $<2.3\times10^{-7}$          & 0.0  \\
\bottomrule
\end{tabular}
\end{table}

The SED therefore indicates that the detected
$\gamma$-ray emission is confined primarily to energies below
approximately
5~GeV,
consistent with the soft spectrum inferred from the global likelihood
fit.

\subsection{Low-Energy Morphology Check (1.0--3.2 GeV)}
\label{sec:low_energy_check}

The broadband spectral energy distribution indicates that the $\gamma$-ray emission associated with PSZ2~G181.06+48.47 is detected almost entirely in the lowest energy interval ($1.0$--$3.2$~GeV), while all higher-energy bins are consistent with upper limits. Motivated by this result, we performed an independent likelihood analysis restricted to the $1.0$--$3.2$~GeV energy range using the same event selection, region of interest, diffuse background components, and FL16Y catalog sources adopted in the broadband analysis.

\begin{figure*}[htbp]
\centering

\begin{minipage}{0.49\textwidth}
    \centering
    \includegraphics[width=\linewidth]{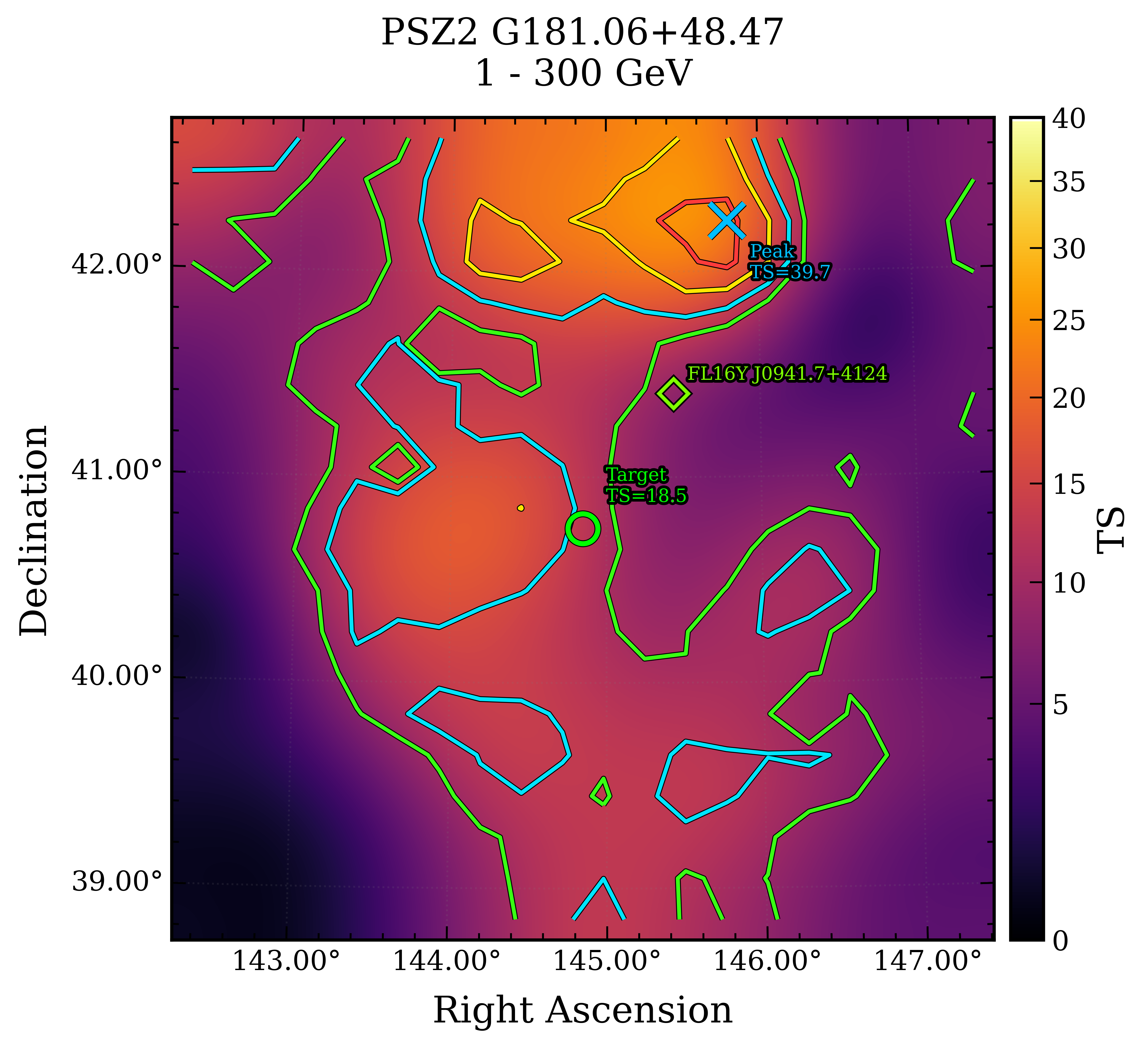}\\[2pt]
    \textbf{(a)} Residual TS map obtained using only the diffuse background and FL16Y catalog sources, without including a source at the position of PSZ2~G181.06+48.47.
\end{minipage}
\hfill
\begin{minipage}{0.49\textwidth}
    \centering
    \includegraphics[width=\linewidth]{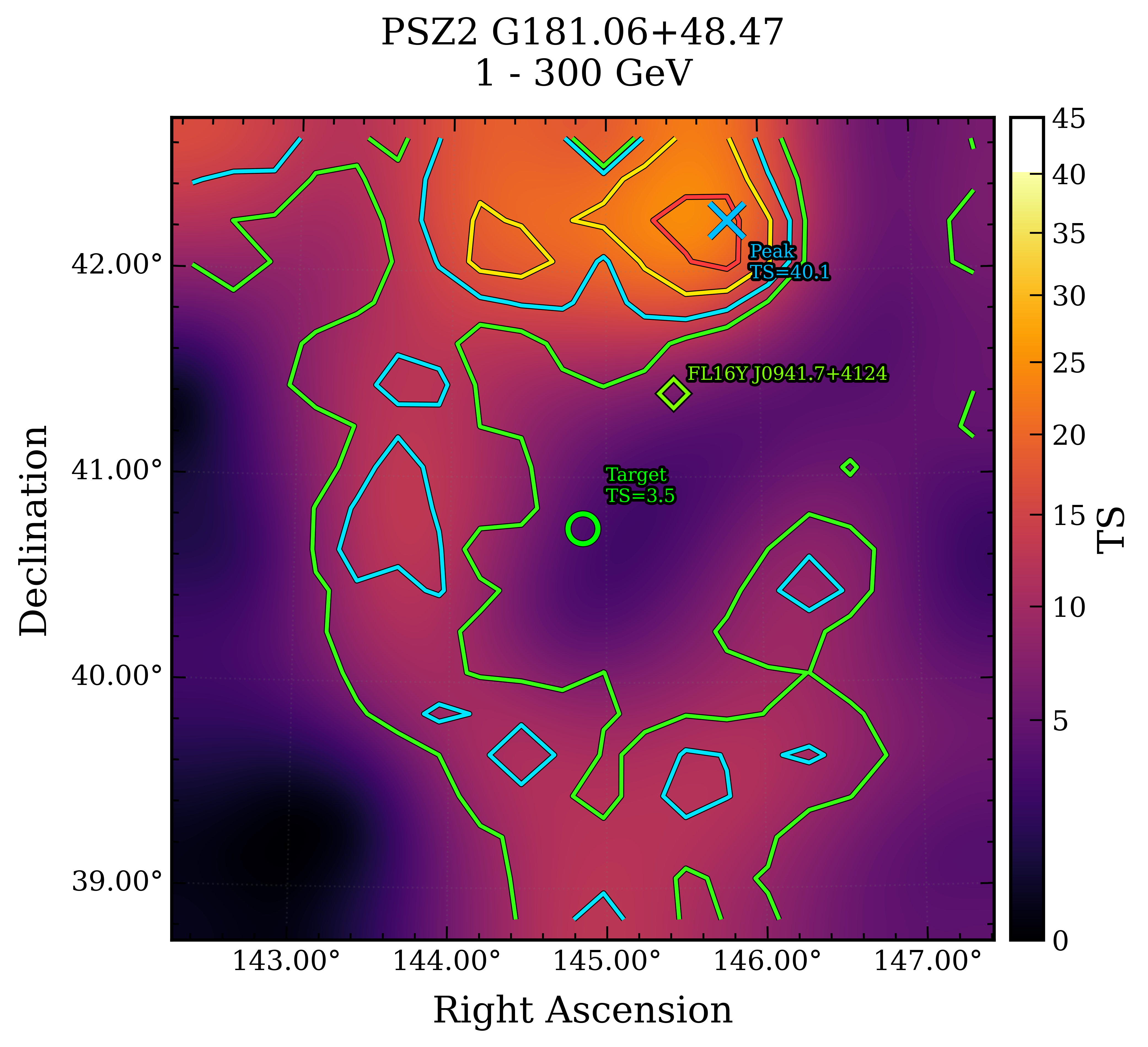}\\[2pt]
    \textbf{(b)} Residual TS map after adding a point source at the cluster position and refitting the model.
\end{minipage}

\caption{
Residual TS maps for the $1.0$--$3.2$~GeV analysis. The left panel shows the residuals obtained using only the diffuse background components and FL16Y catalog sources, revealing a significant excess at the position of PSZ2~G181.06+48.47 with $TS_{\rm residual}=18.5$. The right panel shows the residual TS map after introducing a point source at the cluster position, reducing the residual at the cluster center to $TS_{\rm residual}=3.6$. In both panels, the brightest residual hotspot remains at $(\alpha,\delta)\approx(145.80^{\circ},\,42.25^{\circ})$, demonstrating that it is spatially distinct from the cluster emission.
}
\label{fig:low_energy_tsmap}
\end{figure*}

As a first step, no source was included at the position of the cluster, such that the model consisted only of the Galactic diffuse emission, isotropic background, and catalog sources. The likelihood fit yielded $-\log\mathcal{L}=37636.4$. The corresponding residual TS map (Fig.~\ref{fig:low_energy_tsmap}) shows a significant excess at the cluster position with $TS_{\rm residual}=18.5$, while the brightest residual hotspot has $TS_{\rm max}=39.69$ and is located at $(\alpha,\delta)=(145.80^{\circ},\,42.25^{\circ})$, consistent with the neighboring hotspot identified in the broadband analysis.

A point source with a PowerLaw spectrum was then introduced at the nominal cluster position $(\alpha,\delta)=(144.85^{\circ},\,40.75^{\circ})$, allowing both the normalization and photon index to vary during the likelihood fit. The best-fitting model yields a source detection significance of $TS=15.3$, a photon index of $\Gamma=3.86\pm1.54$, and an integrated photon flux of $(8.63\pm2.66)\times10^{-11}\,\mathrm{ph\,cm^{-2}\,s^{-1}}$. The likelihood improves to $-\log\mathcal{L}=37629.18$, corresponding to $\Delta TS=14.4$ relative to the background-only model. The residual TS map after introducing the point source is shown in the right panel of Fig.~\ref{fig:low_energy_tsmap}. The residual emission at the cluster position is reduced to $TS_{\rm residual}=3.55$, whereas the neighboring hotspot remains essentially unchanged with $TS_{\rm max}=40.1$ at the same sky position. These results demonstrate that the low-energy excess is largely accounted for by a source located at the position of PSZ2~G181.06+48.47, while the nearby hotspot remains spatially distinct from the cluster emission. The analysis therefore confirms that the broadband detection is dominated by photons below $\sim3$~GeV and that the neighboring excess is an independent feature rather than residual emission from the cluster.

\subsubsection{Simultaneous Modeling of the Cluster and Neighboring Hotspot}
\label{sec:dual_gaussian}

To verify that the extended emission associated with PSZ2~G181.06+48.47 is not influenced by the nearby residual hotspot identified in the broadband analysis, a final likelihood analysis was performed in the $1.0$--$3.2$~GeV energy range by simultaneously modeling both excesses. The cluster was represented by a RadialGaussian spatial template centered at $(\alpha,\delta)=(144.85^{\circ},\,40.75^{\circ})$, while the neighboring hotspot was modeled using a second RadialGaussian centered at $(145.525^{\circ},\,42.248^{\circ})$. For both sources, the Gaussian width was fixed at $\sigma=0.4^{\circ}$, corresponding to the preferred spatial model obtained from the broadband analysis, while the spectral normalization and photon index were allowed to vary during the likelihood fit.

\begin{figure}[htbp]
\centering
\includegraphics[width=0.72\linewidth]{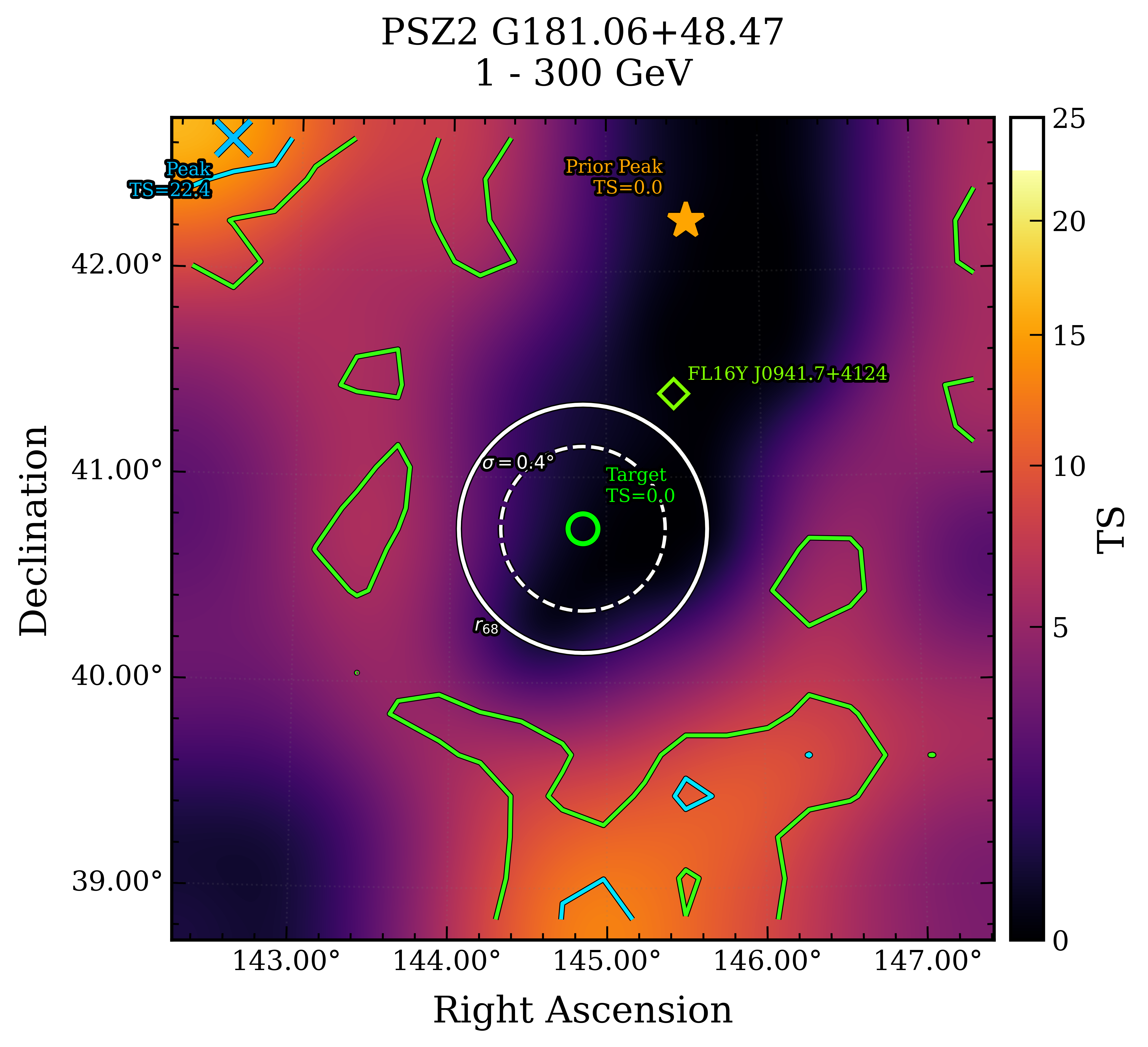}
\caption{
Residual TS map for the $1.0$--$3.2$~GeV analysis after simultaneously modeling PSZ2~G181.06+48.47 and the neighboring hotspot with two RadialGaussian spatial templates of fixed width $\sigma=0.4^{\circ}$. No statistically significant residual emission remains at either source position ($TS_{\rm residual}=0.0$ for both the cluster and the neighboring hotspot). The brightest residual in the field has $TS_{\rm max}=22.4$ and is located approximately $2.55^{\circ}$ from the cluster center, indicating that it is unrelated to the two modeled sources. This provides strong evidence that the two excesses are distinct emission components and that the extended emission associated with PSZ2~G181.06+48.47 is robust against contamination from the neighboring hotspot.
}
\label{fig:dual_gaussian_residual}
\end{figure}

The simultaneous fit yields a photon index of $\Gamma=3.21\pm0.75$ and an integrated photon flux of $(2.66\pm0.46)\times10^{-10}\,\mathrm{ph\,cm^{-2}\,s^{-1}}$ for PSZ2~G181.06+48.47. The neighboring hotspot is described by a photon index of $\Gamma=2.87\pm0.45$ and a photon flux of $(3.22\pm0.45)\times10^{-10}\,\mathrm{ph\,cm^{-2}\,s^{-1}}$. The likelihood improves substantially to $-\log\mathcal{L}=37581.20$, representing the best-fitting model obtained for the restricted $1.0$--$3.2$~GeV analysis.The corresponding residual TS map is shown in Fig.~\ref{fig:dual_gaussian_residual}. After simultaneously modeling both Gaussian components, no statistically significant residual emission remains at either source position, with $TS_{\rm residual}=0.0$ at the cluster center and $TS_{\rm residual}=0.0$ at the neighboring hotspot. The brightest remaining residual in the field has $TS_{\rm max}=22.4$ at $(\alpha,\delta)=(142.54^{\circ},\,42.63^{\circ})$, approximately $2.55^{\circ}$ from the cluster center, indicating that it is unrelated to either of the two modeled sources. This analysis demonstrates that the low-energy $\gamma$-ray emission is well described by two independent extended Gaussian components. Simultaneously modeling both sources completely removes the residual emission at their respective positions, confirming that the extended emission associated with PSZ2~G181.06+48.47 is not an artifact of contamination from the neighboring hotspot.

\section{Conclusions}
\label{sec:conclusions}

We have presented a detailed analysis of the galaxy cluster
PSZ2~G181.06+48.47 using 17.9 years of \textit{Fermi}-LAT Pass~8
SOURCE-class data in the 1--300~GeV energy range. Our principal
findings are summarized below.

\begin{itemize}

\item A significant $\gamma$-ray excess is detected at the position of
PSZ2~G181.06+48.47. A point-source model provides a statistically
significant detection
($TS_{\rm source}=19.0$;
$\Gamma=3.20\pm0.62$),
but leaves a residual excess at the cluster position, indicating that a
point-like morphology does not fully describe the observed emission.

\item Replacing the point source with a
\texttt{RadialGaussian} spatial template significantly improves the
likelihood. The preferred model is obtained for a Gaussian width of
$\sigma=0.4^{\circ}$, yielding a source detection significance of
$TS_{\rm source}=55.0$
($\approx7.4\sigma$)
and an extension test statistic of
$TS_{\rm ext}=36.8$
relative to the point-source hypothesis. The best-fitting model has a
photon index of
$\Gamma=2.62\pm0.26$
and an integrated photon flux of
$(3.22\pm0.50)\times10^{-10}\,
\mathrm{ph\,cm^{-2}\,s^{-1}}$.
The residual TS at the cluster position is reduced to a value
consistent with zero, demonstrating that the extended Gaussian template
provides a substantially better description of the observed emission
than a point-source model.

\item A \texttt{RadialDisk} template also improves the likelihood
relative to a point source, but the likelihood increases monotonically
with disk radius and does not exhibit a well-defined maximum.
Furthermore, sufficiently large disk templates begin to absorb unrelated
emission elsewhere in the ROI. We therefore conclude that the
\texttt{RadialDisk} model does not provide a physically meaningful
estimate of the intrinsic extent of the cluster emission.

\item A second significant $\gamma$-ray excess is detected
approximately $1.58^{\circ}$ north of the cluster, spatially coincident
with 4FGL~J0943.6+4207. Simultaneous modelling of both excesses removes
the residual emission at their respective positions and demonstrates
that the inferred spectral and spatial properties of
PSZ2~G181.06+48.47 are robust against contamination from this
neighbouring source. Within the adopted ROI and background model, the
neighbouring hotspot is better described by a Gaussian template than by
a point source, although this should not be interpreted as a formal
extension measurement of 4FGL~J0943.6+4207 itself.

\item The spectral energy distribution is significantly detected only in
the lowest energy interval
(1.0--3.13~GeV), with all higher-energy bins yielding only upper
limits. An independent likelihood analysis restricted to the
1.0--3.2~GeV energy range reproduces the same spatial behaviour observed
in the broadband analysis, confirming that the detected emission is
dominated by low-energy photons and that the extended morphology is
robust.

\end{itemize}

Taken together, these results provide strong evidence that the
$\gamma$-ray emission associated with PSZ2~G181.06+48.47 is spatially
extended, with a preferred Gaussian width of
$\sigma\approx0.4^{\circ}$
($r_{68}\approx0.60^{\circ}$), and cannot be adequately described by a
point-source morphology. The preferred extended model remains stable
after explicitly accounting for the neighbouring
$\gamma$-ray hotspot, indicating that the inferred extension is
intrinsic to the cluster rather than an artifact of source confusion.
The soft spectrum, together with the confinement of the detected signal
to energies below approximately 5~GeV, is consistent with diffuse
high-energy emission associated with the intracluster medium, although
the underlying physical emission mechanism cannot be uniquely determined
from the present data alone.

As a natural extension of this work, we plan to model the expected
non-thermal emission from PSZ2~G181.06+48.47 using the
\texttt{MINOT} framework, following the methodology of
Manna \& Desai~\cite{Manna2026}. By computing theoretical
$\gamma$-ray spectra and fluxes for different cosmic-ray populations
and intracluster medium models, these calculations will enable a direct
comparison between physically motivated predictions and the
\textit{Fermi}-LAT observations presented here, thereby providing
stronger constraints on the origin of the detected emission and on
particle acceleration processes in galaxy clusters.
\begin{acknowledgments}
AD has been supported by a Summer Undergraduate Research Exposure (SURE) Internship at IIT Hyderabad.SM gratefully acknowledges the Ministry of Education (MoE), Government of India, for its consistent financial support through the research fellowship, which has been instrumental in facilitating the successful completion of this work. 

\end{acknowledgments}

\bibliography{references}

@ARTICLE{Kravtsov2012,
       author = {{Kravtsov}, Andrey V. and {Borgani}, Stefano},
        title = "{Formation of Galaxy Clusters}",
      journal = {\araa},
         year = 2012,
        month = sep,
       volume = {50},
        pages = {353-409},
          doi = {10.1146/annurev-astro-081811-125502},
archivePrefix = {arXiv},
       eprint = {1205.5556},
 primaryClass = {astro-ph.CO},
       adsurl = {https://ui.adsabs.harvard.edu/abs/2012ARA&A..50..353K}
}

@INPROCEEDINGS{Fermipy,
       author = {{Wood}, M. and {Caputo}, R. and {Charles}, E. and {Di Mauro}, M. and {Magill}, J. and {Perkins}, J.~S. and {Fermi-LAT Collaboration}},
        title = "{Fermipy: An open-source Python package for analysis of Fermi-LAT Data}",
    booktitle = {35th International Cosmic Ray Conference (ICRC2017)},
         year = 2017,
       series = {International Cosmic Ray Conference},
       volume = {301},
        month = jul,
          eid = {824},
        pages = {824},
          doi = {10.22323/1.301.0824},
archivePrefix = {arXiv},
       eprint = {1707.09551},
 primaryClass = {astro-ph.IM},
       adsurl = {https://ui.adsabs.harvard.edu/abs/2017ICRC...35..824W}
}

@ARTICLE{Allen2011,
       author = {{Allen}, Steven W. and {Evrard}, August E. and {Mantz}, Adam B.},
        title = "{Cosmological Parameters from Observations of Galaxy Clusters}",
      journal = {\araa},
         year = 2011,
        month = sep,
       volume = {49},
       number = {1},
        pages = {409-470},
          doi = {10.1146/annurev-astro-081710-102514},
archivePrefix = {arXiv},
       eprint = {1103.4829},
 primaryClass = {astro-ph.CO},
       adsurl = {https://ui.adsabs.harvard.edu/abs/2011ARA&A..49..409A}
}

@ARTICLE{Desai2018,
       author = {{Desai}, Shantanu},
        title = "{Limit on graviton mass from galaxy cluster Abell 1689}",
      journal = {Physics Letters B},
         year = 2018,
        month = feb,
       volume = {778},
        pages = {325-331},
          doi = {10.1016/j.physletb.2018.01.052},
archivePrefix = {arXiv},
       eprint = {1708.06502},
 primaryClass = {astro-ph.CO},
       adsurl = {https://ui.adsabs.harvard.edu/abs/2018PhLB..778..325D}
}

@ARTICLE{Bora2021,
       author = {{Bora}, Kamal and {Desai}, Shantanu},
        title = "{Constraints on the variation of fine structure constant from joint SPT-SZ and XMM-Newton observations}",
      journal = {\jcap},
         year = 2021,
        month = feb,
       volume = {2021},
       number = {2},
          eid = {012},
        pages = {012},
          doi = {10.1088/1475-7516/2021/02/012},
archivePrefix = {arXiv},
       eprint = {2008.10541},
 primaryClass = {astro-ph.CO},
       adsurl = {https://ui.adsabs.harvard.edu/abs/2021JCAP...02..012B}
}

@ARTICLE{Vikhlininrev2014,
   author = {{Vikhlinin}, A.~A. and {Kravtsov}, A.~V. and {Markevich}, M.~L. and 
   {Sunyaev}, R.~A. and {Churazov}, E.~M.},
    title = "{Clusters of galaxies}",
  journal = {Physics Uspekhi},
     year = 2014,
    month = apr,
   volume = 57,
      eid = {317-341},
    pages = {317-341},
      doi = {10.3367/UFNe.0184.201404a.0339},
   adsurl = {http://adsabs.harvard.edu/abs/2014PhyU...57..317V}
}

@ARTICLE{Bohringer2016,
       author = {{B{\"o}hringer}, Hans and {Chon}, Gayoung},
        title = "{Constraints on neutrino masses from the study of the nearby large-scale structure and galaxy cluster counts}",
      journal = {Modern Physics Letters A},
         year = 2016,
        month = jul,
       volume = {31},
       number = {21},
          eid = {1640008},
        pages = {1640008},
          doi = {10.1142/S0217732316400083},
archivePrefix = {arXiv},
       eprint = {1610.02855},
 primaryClass = {astro-ph.CO},
       adsurl = {https://ui.adsabs.harvard.edu/abs/2016MPLA...3140008B}
}

@ARTICLE{Miyatake25,
       author = {{Miyatake}, Hironao},
        title = "{Cosmology with Galaxy Clusters}",
      journal = {arXiv e-prints},
         year = 2025,
        month = may,
          eid = {arXiv:2505.07697},
        pages = {arXiv:2505.07697},
          doi = {10.48550/arXiv.2505.07697},
archivePrefix = {arXiv},
       eprint = {2505.07697},
 primaryClass = {astro-ph.CO},
       adsurl = {https://ui.adsabs.harvard.edu/abs/2025arXiv250507697M}
}

@ARTICLE{Wik2014,
       author = {{Wik}, Daniel R. and {Hornstrup}, A. and {Molendi}, S. and {Madejski}, G. and {Harrison}, F.~A. and {Zoglauer}, A. and {Grefenstette}, B.~W. and {Gastaldello}, F. and {Madsen}, K.~K. and {Westergaard}, N.~J. and {Ferreira}, D.~D.~M. and {Kitaguchi}, T. and {Pedersen}, K. and {Boggs}, S.~E. and {Christensen}, F.~E. and {Craig}, W.~W. and {Hailey}, C.~J. and {Stern}, D. and {Zhang}, W.~W.},
        title = "{NuSTAR Observations of the Bullet Cluster: Constraints on Inverse Compton Emission}",
      journal = {\apj},
         year = 2014,
        month = sep,
       volume = {792},
       number = {1},
          eid = {48},
        pages = {48},
          doi = {10.1088/0004-637X/792/1/48},
archivePrefix = {arXiv},
       eprint = {1403.2722},
 primaryClass = {astro-ph.HE},
       adsurl = {https://ui.adsabs.harvard.edu/abs/2014ApJ...792...48W}
}

@ARTICLE{Feretti2012,
       author = {{Feretti}, Luigina and {Giovannini}, Gabriele and {Govoni}, Federica and {Murgia}, Matteo},
        title = "{Clusters of galaxies: observational properties of the diffuse radio emission}",
      journal = {\aapr},
         year = 2012,
        month = may,
       volume = {20},
          eid = {54},
        pages = {54},
          doi = {10.1007/s00159-012-0054-z},
archivePrefix = {arXiv},
       eprint = {1205.1919},
 primaryClass = {astro-ph.CO},
       adsurl = {https://ui.adsabs.harvard.edu/abs/2012A&ARv..20...54F}
}

@ARTICLE{Mattox96,
       author = {{Mattox}, J.~R. and {Bertsch}, D.~L. and {Chiang}, J. and {Dingus}, B.~L. and {Digel}, S.~W. and {Esposito}, J.~A. and {Fierro}, J.~M. and {Hartman}, R.~C. and {Hunter}, S.~D. and {Kanbach}, G. and {Kniffen}, D.~A. and {Lin}, Y.~C. and {Macomb}, D.~J. and {Mayer-Hasselwander}, H.~A. and {Michelson}, P.~F. and {von Montigny}, C. and {Mukherjee}, R. and {Nolan}, P.~L. and {Ramanamurthy}, P.~V. and {Schneid}, E. and {Sreekumar}, P. and {Thompson}, D.~J. and {Willis}, T.~D.},
        title = "{The Likelihood Analysis of EGRET Data}",
      journal = {\apj},
         year = 1996,
        month = apr,
       volume = {461},
        pages = {396},
          doi = {10.1086/177068},
       adsurl = {https://ui.adsabs.harvard.edu/abs/1996ApJ...461..396M}
}

@ARTICLE{Ackermann2010,
       author = {{Ackermann}, M. and {Ajello}, M. and {Allafort}, A. and {Baldini}, L. and {Ballet}, J. and {Barbiellini}, G. and {Bastieri}, D. and {Bechtol}, K. and {Bellazzini}, R. and {Blandford}, R.~D. and {Bloom}, E.~D. and {Bonamente}, E. and {Borgland}, A.~W. and {Bouvier}, A. and {Brandt}, T.~J. and {Bregeon}, J. and {Brigida}, M. and {Bruel}, P. and {Buehler}, R. and {Buson}, S. and {Caliandro}, G.~A. and {Cameron}, R.~A. and {Caraveo}, P.~A. and {Carrigan}, S. and {Casandjian}, J.~M. and {Cecchi}, C. and {Charles}, E. and {Chekhtman}, A. and {Cheung}, C.~C. and {Chiang}, J. and {Ciprini}, S. and {Claus}, R. and {Cohen-Tanugi}, J. and {Cominsky}, L.~R. and {Conrad}, J. and {de Angelis}, A. and {de Palma}, F. and {Silva}, E. do Couto e. and {Drell}, P.~S. and {Drlica-Wagner}, A. and {Dubois}, R. and {Dumora}, D. and {Edmonds}, Y. and {Farnier}, C. and {Favuzzi}, C. and {Fegan}, S.~J. and {Frailis}, M. and {Fukazawa}, Y. and {Fusco}, P. and {Gargano}, F. and {Gasparrini}, D. and {Gehrels}, N. and {Germani}, S. and {Giglietto}, N. and {Giordano}, F. and {Glanzman}, T. and {Godfrey}, G. and {Grenier}, I.~A. and {Guiriec}, S. and {Gustafsson}, M. and {Harding}, A.~K. and {Hayashida}, M. and {Horan}, D. and {Hughes}, R.~E. and {Jeltema}, T.~E. and {J{\'o}hannesson}, G. and {Johnson}, A.~S. and {Johnson}, W.~N. and {Kamae}, T. and {Katagiri}, H. and {Kataoka}, J. and {Kn{\"o}dlseder}, J. and {Kuss}, M. and {Lande}, J. and {Latronico}, L. and {Lee}, S. -H. and {Llena Garde}, M. and {Longo}, F. and {Loparco}, F. and {Lovellette}, M.~N. and {Lubrano}, P. and {Madejski}, G.~M. and {Makeev}, A. and {Mazziotta}, M.~N. and {Michelson}, P.~F. and {Mitthumsiri}, W. and {Mizuno}, T. and {Moiseev}, A.~A. and {Monte}, C. and {Monzani}, M.~E. and {Morselli}, A. and {Moskalenko}, I.~V. and {Murgia}, S. and {Nolan}, P.~L. and {Norris}, J.~P. and {Nuss}, E. and {Ohno}, M. and {Ohsugi}, T. and {Omodei}, N. and {Orlando}, E. and {Ormes}, J.~F. and {Panetta}, J.~H. and {Pepe}, M. and {Pesce-Rollins}, M. and {Piron}, F. and {Porter}, T.~A. and {Profumo}, S. and {Rain{\`o}}, S. and {Razzano}, M. and {Reposeur}, T. and {Ritz}, S. and {Rodriguez}, A.~Y. and {Roth}, M. and {Sadrozinski}, H.~F. -W. and {Sander}, A. and {Scargle}, J.~D. and {Sgr{\`o}}, C. and {Siskind}, E.~J. and {Smith}, P.~D. and {Spandre}, G. and {Spinelli}, P. and {Starck}, J. -L. and {Strickman}, M.~S. and {Suson}, D.~J. and {Takahashi}, H. and {Tanaka}, T. and {Thayer}, J.~B. and {Thayer}, J.~G. and {Tibaldo}, L. and {Torres}, D.~F. and {Tosti}, G. and {Usher}, T.~L. and {Vasileiou}, V. and {Vitale}, V. and {Waite}, A.~P. and {Wang}, P. and {Winer}, B.~L. and {Wood}, K.~S. and {Yang}, Z. and {Ylinen}, T. and {Ziegler}, M. and {Fermi LAT Collaboration}},
        title = "{Constraints on dark matter annihilation in clusters of galaxies with the Fermi large area telescope}",
      journal = {\jcap},
         year = 2010,
        month = may,
       volume = {2010},
       number = {5},
          eid = {025},
        pages = {025},
          doi = {10.1088/1475-7516/2010/05/025},
archivePrefix = {arXiv},
       eprint = {1002.2239},
 primaryClass = {astro-ph.CO},
       adsurl = {https://ui.adsabs.harvard.edu/abs/2010JCAP...05..025A}
}

@ARTICLE{Ackermann2014,
       author = {{Ackermann}, M. and {Ajello}, M. and {Albert}, A. and {Allafort}, A. and {Atwood}, W.~B. and {Baldini}, L. and {Ballet}, J. and {Barbiellini}, G. and {Bastieri}, D. and {Bechtol}, K. and {Bellazzini}, R. and {Bloom}, E.~D. and {Bonamente}, E. and {Bottacini}, E. and {Brandt}, T.~J. and {Bregeon}, J. and {Brigida}, M. and {Bruel}, P. and {Buehler}, R. and {Buson}, S. and {Caliandro}, G.~A. and {Cameron}, R.~A. and {Caraveo}, P.~A. and {Cavazzuti}, E. and {Chaves}, R.~C.~G. and {Chiang}, J. and {Chiaro}, G. and {Ciprini}, S. and {Claus}, R. and {Cohen-Tanugi}, J. and {Conrad}, J. and {D'Ammando}, F. and {de Angelis}, A. and {de Palma}, F. and {Dermer}, C.~D. and {Digel}, S.~W. and {Drell}, P.~S. and {Drlica-Wagner}, A. and {Favuzzi}, C. and {Franckowiak}, A. and {Funk}, S. and {Fusco}, P. and {Gargano}, F. and {Gasparrini}, D. and {Germani}, S. and {Giglietto}, N. and {Giordano}, F. and {Giroletti}, M. and {Godfrey}, G. and {Gomez-Vargas}, G.~A. and {Grenier}, I.~A. and {Guiriec}, S. and {Gustafsson}, M. and {Hadasch}, D. and {Hayashida}, M. and {Hewitt}, J. and {Hughes}, R.~E. and {Jeltema}, T.~E. and {J{\'o}hannesson}, G. and {Johnson}, A.~S. and {Kamae}, T. and {Kataoka}, J. and {Kn{\"o}dlseder}, J. and {Kuss}, M. and {Lande}, J. and {Larsson}, S. and {Latronico}, L. and {Llena Garde}, M. and {Longo}, F. and {Loparco}, F. and {Lovellette}, M.~N. and {Lubrano}, P. and {Mayer}, M. and {Mazziotta}, M.~N. and {McEnery}, J.~E. and {Michelson}, P.~F. and {Mitthumsiri}, W. and {Mizuno}, T. and {Monzani}, M.~E. and {Morselli}, A. and {Moskalenko}, I.~V. and {Murgia}, S. and {Nemmen}, R. and {Nuss}, E. and {Ohsugi}, T. and {Orienti}, M. and {Orlando}, E. and {Ormes}, J.~F. and {Perkins}, J.~S. and {Pesce-Rollins}, M. and {Piron}, F. and {Pivato}, G. and {Rain{\`o}}, S. and {Rando}, R. and {Razzano}, M. and {Razzaque}, S. and {Reimer}, A. and {Reimer}, O. and {Ruan}, J. and {S{\'a}nchez-Conde}, M. and {Schulz}, A. and {Sgr{\`o}}, C. and {Siskind}, E.~J. and {Spandre}, G. and {Spinelli}, P. and {Storm}, E. and {Strong}, A.~W. and {Suson}, D.~J. and {Takahashi}, H. and {Thayer}, J.~G. and {Thayer}, J.~B. and {Thompson}, D.~J. and {Tibaldo}, L. and {Tinivella}, M. and {Torres}, D.~F. and {Troja}, E. and {Uchiyama}, Y. and {Usher}, T.~L. and {Vandenbroucke}, J. and {Vianello}, G. and {Vitale}, V. and {Winer}, B.~L. and {Wood}, K.~S. and {Zimmer}, S. and {Fermi-LAT Collaboration} and {Pinzke}, A. and {Pfrommer}, C.},
        title = "{Search for Cosmic-Ray-induced Gamma-Ray Emission in Galaxy Clusters}",
      journal = {\apj},
         year = 2014,
        month = may,
       volume = {787},
       number = {1},
          eid = {18},
        pages = {18},
          doi = {10.1088/0004-637X/787/1/18},
archivePrefix = {arXiv},
       eprint = {1308.5654},
 primaryClass = {astro-ph.HE},
       adsurl = {https://ui.adsabs.harvard.edu/abs/2014ApJ...787...18A}
}

@ARTICLE{Atwood09,
       author = {{Atwood}, W.~B. and {Abdo}, A.~A. and {Ackermann}, M. and {Althouse}, W. and {Anderson}, B. and {Axelsson}, M. and {Baldini}, L. and {Ballet}, J. and {Band}, D.~L. and {Barbiellini}, G. and {Bartelt}, J. and others},
        title = "{The Large Area Telescope on the Fermi Gamma-Ray Space Telescope Mission}",
      journal = {\apj},
         year = 2009,
        month = jun,
       volume = {697},
       number = {2},
        pages = {1071-1102},
          doi = {10.1088/0004-637X/697/2/1071},
archivePrefix = {arXiv},
       eprint = {0902.1089},
 primaryClass = {astro-ph.IM},
       adsurl = {https://ui.adsabs.harvard.edu/abs/2009ApJ...697.1071A}
}

@ARTICLE{Ackermann2016,
       author = {{Ackermann}, M. and {Albert}, A. and {Atwood}, W.~B. and {Baldini}, L. and {Ballet}, J. and {Barbiellini}, G. and {Bastieri}, D. and {Bellazzini}, R. and {Bissaldi}, E. and {Bloom}, E.~D. and {Bonino}, R. and {Brandt}, T.~J. and {Bregeon}, J. and {Bruel}, P. and {Buehler}, R. and {Caliandro}, G.~A. and {Cameron}, R.~A. and {Caragiulo}, M. and {Caraveo}, P.~A. and {Cavazzuti}, E. and {Cecchi}, C. and {Charles}, E. and {Chekhtman}, A. and {Chiang}, J. and {Chiaro}, G. and {Ciprini}, S. and {Cohen-Tanugi}, J. and {Cutini}, S. and {D'Ammando}, F. and {de Angelis}, A. and {de Palma}, F. and {Desiante}, R. and {Digel}, S.~W. and {Drell}, P.~S. and {Favuzzi}, C. and {Ferrara}, E.~C. and {Focke}, W.~B. and {Franckowiak}, A. and {Fusco}, P. and {Gargano}, F. and {Gasparrini}, D. and {Giglietto}, N. and {Giordano}, F. and {Godfrey}, G. and {Grenier}, I.~A. and {Grondin}, M.-H. and {Guillemot}, L. and {Guiriec}, S. and {Harding}, A.~K. and {Hill}, A.~B. and {Horan}, D. and {J{\'o}hannesson}, G. and {Kn{\"o}dlseder}, J. and {Kuss}, M. and {Larsson}, S. and {Latronico}, L. and {Li}, J. and {Li}, L. and {Longo}, F. and {Loparco}, F. and {Lubrano}, P. and {Maldera}, S. and {Martin}, P. and {Mayer}, M. and {Mazziotta}, M.~N. and {Michelson}, P.~F. and {Mizuno}, T. and {Monzani}, M.~E. and {Morselli}, A. and {Murgia}, S. and {Nuss}, E. and {Ohsugi}, T. and {Orienti}, M. and {Orlando}, E. and {Ormes}, J.~F. and {Paneque}, D. and {Pesce-Rollins}, M. and {Piron}, F. and {Pivato}, G. and {Porter}, T.~A. and {Rain{\`o}}, S. and {Rando}, R. and {Razzano}, M. and {Reimer}, A. and {Reimer}, O. and {Romani}, R.~W. and {S{\'a}nchez-Conde}, M. and {Schulz}, A. and {Sgr{\`o}}, C. and {Siskind}, E.~J. and {Smith}, D.~A. and {Spada}, F. and {Spandre}, G. and {Spinelli}, P. and {Suson}, D.~J. and {Takahashi}, H. and {Thayer}, J.~B. and {Tibaldo}, L. and {Torres}, D.~F. and {Tosti}, G. and {Troja}, E. and {Vianello}, G. and {Wood}, M. and {Zimmer}, S.},
        title = "{Deep view of the Large Magellanic Cloud with six years of Fermi-LAT observations}",
      journal = {\aap},
         year = 2016,
        month = feb,
       volume = {586},
          eid = {A71},
        pages = {A71},
          doi = {10.1051/0004-6361/201526920},
archivePrefix = {arXiv},
       eprint = {1509.06903},
 primaryClass = {astro-ph.HE},
       adsurl = {https://ui.adsabs.harvard.edu/abs/2016A&A...586A..71A}
}

@ARTICLE{Fermilat2026,
       author = {{Ballet}, J. and {Bruel}, P. and {Burnett}, T.~H. and {Lott}, B.},
        title = "{Fermi-LAT 16-year Source List}",
      journal = {arXiv e-prints},
         year = 2026,
        month = feb,
          eid = {arXiv:2602.22148},
        pages = {arXiv:2602.22148},
          doi = {10.48550/arXiv.2602.22148},
archivePrefix = {arXiv},
       eprint = {2602.22148},
 primaryClass = {astro-ph.HE},
       adsurl = {https://ui.adsabs.harvard.edu/abs/2026arXiv260222148B}
}

@ARTICLE{Harale2025,
       author = {{Harale}, Gajanan D. and {Paul}, Surajit},
        title = "{Excess of diffuse gamma-ray emission detected from the galaxy cluster Abell 119 from 14-year Fermi-LAT data}",
      journal = {\prd},
         year = 2025,
        month = nov,
       volume = {112},
       number = {10},
          eid = {103025},
        pages = {103025},
          doi = {10.1103/gn1q-pzx3},
archivePrefix = {arXiv},
       eprint = {2511.15559},
 primaryClass = {astro-ph.HE},
       adsurl = {https://ui.adsabs.harvard.edu/abs/2025PhRvD.112j3025H}
}

@ARTICLE{Keshet2025a,
       author = {{Keshet}, Uri},
        title = "{Galaxy-cluster-stacked Fermi-LAT. Part II. Extended central hadronic signal}",
      journal = {\jcap},
         year = 2025,
        month = oct,
       volume = {2025},
       number = {10},
          eid = {016},
        pages = {016},
          doi = {10.1088/1475-7516/2025/10/016},
archivePrefix = {arXiv},
       eprint = {2502.19494},
 primaryClass = {astro-ph.HE},
       adsurl = {https://ui.adsabs.harvard.edu/abs/2025JCAP...10..016K}
}

@ARTICLE{ReissKeshet2018,
       author = {{Reiss}, Ido and {Keshet}, Uri},
        title = "{Detection of virial shocks in stacked Fermi-LAT galaxy clusters}",
      journal = {\jcap},
         year = 2018,
        month = oct,
       volume = {2018},
       number = {10},
          eid = {010},
        pages = {010},
          doi = {10.1088/1475-7516/2018/10/010},
archivePrefix = {arXiv},
       eprint = {1705.05376},
 primaryClass = {astro-ph.HE},
       adsurl = {https://ui.adsabs.harvard.edu/abs/2018JCAP...10..010R}
}

@ARTICLE{Baghmanyan2022,
       author = {{Baghmanyan}, Vardan and {Zargaryan}, Davit and {Aharonian}, Felix and {Yang}, Ruizhi and {Casanova}, Sabrina and {Mackey}, Jonathan},
        title = "{Detailed study of extended {\ensuremath{\gamma}}-ray morphology in the vicinity of the Coma cluster with Fermi Large Area Telescope}",
      journal = {\mnras},
         year = 2022,
        month = oct,
       volume = {516},
       number = {1},
        pages = {562-571},
          doi = {10.1093/mnras/stac2266},
archivePrefix = {arXiv},
       eprint = {2110.00309},
 primaryClass = {astro-ph.HE},
       adsurl = {https://ui.adsabs.harvard.edu/abs/2022MNRAS.516..562B}
}

@ARTICLE{Mannastacked,
       author = {{Manna}, Siddhant and {Desai}, Shantanu},
        title = "{A stacked analysis of GeV gamma-ray emission from SPT-SZ galaxy clusters with 16 years of Fermi-LAT data}",
      journal = {Physics of the Dark Universe},
         year = 2025,
        month = sep,
       volume = {49},
          eid = {101966},
        pages = {101966},
          doi = {10.1016/j.dark.2025.101966},
archivePrefix = {arXiv},
       eprint = {2502.15235},
 primaryClass = {astro-ph.HE},
       adsurl = {https://ui.adsabs.harvard.edu/abs/2025PDU....4901966M}
}

@ARTICLE{ShangLi,
       author = {{Li}, Shang and {Han}, Feng},
        title = "{Search for {\ensuremath{\gamma}}-Ray Emission from Cluster of Galaxies with Fermi-LAT Data}",
      journal = {\apj},
         year = 2026,
        month = feb,
       volume = {997},
       number = {2},
          eid = {227},
        pages = {227},
          doi = {10.3847/1538-4357/ae2bda},
       adsurl = {https://ui.adsabs.harvard.edu/abs/2026ApJ...997..227L}
}

@ARTICLE{Manna2024,
       author = {{Manna}, Siddhant and {Desai}, Shantanu},
        title = "{Search for GeV gamma-ray emission from SPT-SZ selected galaxy clusters with 15 years of Fermi-LAT data}",
      journal = {\jcap},
         year = 2024,
        month = jan,
       volume = {2024},
       number = {1},
          eid = {017},
        pages = {017},
          doi = {10.1088/1475-7516/2024/01/017},
archivePrefix = {arXiv},
       eprint = {2310.07519},
 primaryClass = {astro-ph.HE},
       adsurl = {https://ui.adsabs.harvard.edu/abs/2024JCAP...01..017M}
}

@ARTICLE{Brunetti14,
       author = {{Brunetti}, Gianfranco and {Jones}, Thomas W.},
        title = "{Cosmic Rays in Galaxy Clusters and Their Nonthermal Emission}",
      journal = {International Journal of Modern Physics D},
         year = 2014,
        month = mar,
       volume = {23},
       number = {4},
          eid = {1430007-98},
        pages = {1430007-98},
          doi = {10.1142/S0218271814300079},
archivePrefix = {arXiv},
       eprint = {1401.7519},
 primaryClass = {astro-ph.CO},
       adsurl = {https://ui.adsabs.harvard.edu/abs/2014IJMPD..2330007B}
}

@ARTICLE{Ahn25,
       author = {{Ahn}, Eunmo and {Cho}, Hyejeon and {Jee}, M. James and {Lee}, Wonki and {Stroe}, Andra and {Rajpurohit}, Kamlesh and {Finner}, Kyle and {Forman}, William and {Jones}, Christine and {van Weeren}, Reinout},
        title = "{PSZ2 G181.06+48.47. III. Weak-lensing Analysis and Merging Scenario Reconstruction of a Low-mass Cluster with Exceptionally Distant Radio Relics}",
      journal = {\apj},
         year = 2025,
        month = may,
       volume = {984},
       number = {1},
          eid = {26},
        pages = {26},
          doi = {10.3847/1538-4357/adbc99},
archivePrefix = {arXiv},
       eprint = {2501.09067},
 primaryClass = {astro-ph.CO},
       adsurl = {https://ui.adsabs.harvard.edu/abs/2025ApJ...984...26A}
}

@ARTICLE{Stroe25,
       author = {{Stroe}, Andra and {Rajpurohit}, Kamlesh and {Zhu}, Zhenlin and {Lovisari}, Lorenzo and {Simionescu}, Aurora and {O'Sullivan}, Ewan and {Randall}, Scott and {Forman}, William and {Akamatsu}, Hiroki and {van Weeren}, Reinout and {Jee}, M. James and {Lee}, Wonki and {Cho}, Hyejeon and {Ahn}, Eunmo and {Finner}, Kyle and {Jones}, Christine},
        title = "{PSZ2 G181.06+48.47. I. X-Ray Exploration of a Low-mass Cluster with Exceptionally Distant Radio Relics}",
      journal = {\apj},
         year = 2025,
        month = may,
       volume = {984},
       number = {1},
          eid = {24},
        pages = {24},
          doi = {10.3847/1538-4357/adb731},
archivePrefix = {arXiv},
       eprint = {2501.07651},
 primaryClass = {astro-ph.HE},
       adsurl = {https://ui.adsabs.harvard.edu/abs/2025ApJ...984...24S}
}

@ARTICLE{Rajpurohit25,
       author = {{Rajpurohit}, Kamlesh and {Stroe}, Andra and {O'Sullivan}, Ewan and {Ahn}, Eunmo and {Lee}, Wonki and {Cho}, Hyejeon and {Jee}, M. James and {van Weeren}, Reinout and {Lovisari}, Lorenzo and {Finner}, Kyle and {Simionescu}, Aurora and {Forman}, William and {Shimwell}, Timothy and {Jones}, Christine and {Zhu}, Zhenlin and {Randall}, Scott},
        title = "{PSZ2 G181.06+48.47. II. Radio Analysis of a Low-mass Cluster with Exceptionally Distant Radio Relics}",
      journal = {\apj},
         year = 2025,
        month = may,
       volume = {984},
       number = {1},
          eid = {25},
        pages = {25},
          doi = {10.3847/1538-4357/adbbb9},
archivePrefix = {arXiv},
       eprint = {2501.08390},
 primaryClass = {astro-ph.CO},
       adsurl = {https://ui.adsabs.harvard.edu/abs/2025ApJ...984...25R}
}

@ARTICLE{Planck16,
       author = {{Planck Collaboration} and {Ade}, P.~A.~R. and {Aghanim}, N. and {Arnaud}, M. and {Ashdown}, M. and {Aumont}, J. and {Baccigalupi}, C. and {Banday}, A.~J. and {Barreiro}, R.~B. and {Bartlett}, J.~G. and {Bartolo}, N. and {Battaner}, E. and {Battye}, R. and {Benabed}, K. and {Beno{\^\i}t}, A. and {Benoit-L{\'e}vy}, A. and {Bernard}, J.-P. and {Bersanelli}, M. and {Bielewicz}, P. and {Bock}, J.~J. and {Bonaldi}, A. and {Bonavera}, L. and {Bond}, J.~R. and {Borrill}, J. and {Bouchet}, F.~R. and {Boulanger}, F. and {Bucher}, M. and {Burigana}, C. and {Butler}, R.~C. and {Calabrese}, E. and {Cardoso}, J.-F. and {Catalano}, A. and {Challinor}, A. and {Chamballu}, A. and {Chary}, R.-R. and {Chiang}, H.~C. and {Chluba}, J. and {Christensen}, P.~R. and {Church}, S. and {Clements}, D.~L. and {Colombi}, S. and {Colombo}, L.~P.~L. and {Combet}, C. and {Coulais}, A. and {Crill}, B.~P. and {Curto}, A. and {Cuttaia}, F. and {Danese}, L. and {Davies}, R.~D. and {Davis}, R.~J. and {de Bernardis}, P. and {de Rosa}, A. and {de Zotti}, G. and {Delabrouille}, J. and {D{\'e}sert}, F.-X. and {Di Valentino}, E. and {Dickinson}, C. and {Diego}, J.~M. and {Dolag}, K. and {Dole}, H. and {Donzelli}, S. and {Dor{\'e}}, O. and {Douspis}, M. and {Ducout}, A. and {Dunkley}, J. and {Dupac}, X. and {Efstathiou}, G. and {Elsner}, F. and {En{\ss}lin}, T.~A. and {Eriksen}, H.~K. and {Farhang}, M. and {Fergusson}, J. and {Finelli}, F. and {Forni}, O. and {Frailis}, M. and {Fraisse}, A.~A. and {Franceschi}, E. and {Frejsel}, A. and {Galeotta}, S. and {Galli}, S. and {Ganga}, K. and {Gauthier}, C. and {Gerbino}, M. and {Ghosh}, T. and {Giard}, M. and {Giraud-H{\'e}raud}, Y. and {Giusarma}, E. and {Gjerl{\o}w}, E. and {Gonz{\'a}lez-Nuevo}, J. and {G{\'o}rski}, K.~M. and {Gratton}, S. and {Gregorio}, A. and {Gruppuso}, A. and {Gudmundsson}, J.~E. and {Hamann}, J. and {Hansen}, F.~K. and {Hanson}, D. and {Harrison}, D.~L. and {Helou}, G. and {Henrot-Versill{\'e}}, S. and {Hern{\'a}ndez-Monteagudo}, C. and {Herranz}, D. and {Hildebrandt}, S.~R. and {Hivon}, E. and {Hobson}, M. and {Holmes}, W.~A. and {Hornstrup}, A. and {Hovest}, W. and {Huang}, Z. and {Huffenberger}, K.~M. and {Hurier}, G. and {Jaffe}, A.~H. and {Jaffe}, T.~R. and {Jones}, W.~C. and {Juvela}, M. and {Keih{\"a}nen}, E. and {Keskitalo}, R. and {Kisner}, T.~S. and {Kneissl}, R. and {Knoche}, J. and {Knox}, L. and {Kunz}, M. and {Kurki-Suonio}, H. and {Lagache}, G. and {L{\"a}hteenm{\"a}ki}, A. and {Lamarre}, J.-M. and {Lasenby}, A. and {Lattanzi}, M. and {Lawrence}, C.~R. and {Leahy}, J.~P. and {Leonardi}, R. and {Lesgourgues}, J. and {Levrier}, F. and {Lewis}, A. and {Liguori}, M. and {Lilje}, P.~B. and {Linden-V{\o}rnle}, M. and {L{\'o}pez-Caniego}, M. and {Lubin}, P.~M. and {Mac{\'\i}as-P{\'e}rez}, J.~F. and {Maggio}, G. and {Maino}, D. and {Mandolesi}, N. and {Mangilli}, A. and {Marchini}, A. and {Maris}, M. and {Martin}, P.~G. and {Martinelli}, M. and {Mart{\'\i}nez-Gonz{\'a}lez}, E. and {Masi}, S. and {Matarrese}, S. and {McGehee}, P. and {Meinhold}, P.~R. and {Melchiorri}, A. and {Melin}, J.-B. and {Mendes}, L. and {Mennella}, A. and {Migliaccio}, M. and {Millea}, M. and {Mitra}, S. and {Miville-Desch{\^e}nes}, M.-A. and {Moneti}, A. and {Montier}, L. and {Morgante}, G. and {Mortlock}, D. and {Moss}, A. and {Munshi}, D. and {Murphy}, J.~A. and {Naselsky}, P. and {Nati}, F. and {Natoli}, P. and {Netterfield}, C.~B. and {N{\o}rgaard-Nielsen}, H.~U. and {Noviello}, F. and {Novikov}, D. and {Novikov}, I. and {Oxborrow}, C.~A. and {Paci}, F. and {Pagano}, L. and {Pajot}, F. and {Paladini}, R. and {Paoletti}, D. and {Partridge}, B. and {Pasian}, F. and {Patanchon}, G. and {Pearson}, T.~J. and {Perdereau}, O. and {Perotto}, L. and {Perrotta}, F. and {Pettorino}, V. and {Piacentini}, F. and {Piat}, M. and {Pierpaoli}, E. and {Pietrobon}, D. and {Plaszczynski}, S. and {Pointecouteau}, E. and {Polenta}, G. and {Popa}, L. and {Pratt}, G.~W. and {Pr{\'e}zeau}, G.},
        title = "{Planck 2015 results. XIII. Cosmological parameters}",
      journal = {\aap},
         year = 2016,
        month = sep,
       volume = {594},
          eid = {A13},
        pages = {A13},
          doi = {10.1051/0004-6361/201525830},
archivePrefix = {arXiv},
       eprint = {1502.01589},
 primaryClass = {astro-ph.CO},
       adsurl = {https://ui.adsabs.harvard.edu/abs/2016A&A...594A..13P}
}

@ARTICLE{Rykoff,
       author = {{Rykoff}, E.~S. and {Rozo}, E. and {Busha}, M.~T. and {Cunha}, C.~E. and {Finoguenov}, A. and {Evrard}, A. and {Hao}, J. and {Koester}, B.~P. and {Leauthaud}, A. and {Nord}, B. and {Pierre}, M. and {Reddick}, R. and {Sadibekova}, T. and {Sheldon}, E.~S. and {Wechsler}, R.~H.},
        title = "{redMaPPer. I. Algorithm and SDSS DR8 Catalog}",
      journal = {\apj},
         year = 2014,
        month = apr,
       volume = {785},
       number = {2},
          eid = {104},
        pages = {104},
          doi = {10.1088/0004-637X/785/2/104},
archivePrefix = {arXiv},
       eprint = {1303.3562},
 primaryClass = {astro-ph.CO},
       adsurl = {https://ui.adsabs.harvard.edu/abs/2014ApJ...785..104R}
}

@ARTICLE{Jones23,
       author = {{Jones}, A. and {de Gasperin}, F. and {Cuciti}, V. and {Botteon}, A. and {Zhang}, X. and {Gastaldello}, F. and {Shimwell}, T. and {Simionescu}, A. and {Rossetti}, M. and {Cassano}, R. and {Akamatsu}, H. and {Bonafede}, A. and {Br{\"u}ggen}, M. and {Brunetti}, G. and {Camillini}, L. and {Di Gennaro}, G. and {Drabent}, A. and {Hoang}, D.~N. and {Rajpurohit}, K. and {Natale}, R. and {Tasse}, C. and {van Weeren}, R.~J.},
        title = "{The Planck clusters in the LOFAR sky. VI. LoTSS-DR2: Properties of radio relics}",
      journal = {\aap},
         year = 2023,
        month = dec,
       volume = {680},
          eid = {A31},
        pages = {A31},
          doi = {10.1051/0004-6361/202245102},
archivePrefix = {arXiv},
       eprint = {2301.07814},
 primaryClass = {astro-ph.CO},
       adsurl = {https://ui.adsabs.harvard.edu/abs/2023A&A...680A..31J}
}

@ARTICLE{4FGL-DR4,
       author = {{Ballet}, J. and {Bruel}, P. and {Burnett}, T.~H. and {Lott}, B. and {The Fermi-LAT collaboration}},
        title = "{Fermi Large Area Telescope Fourth Source Catalog Data Release 4 (4FGL-DR4)}",
      journal = {arXiv e-prints},
         year = 2023,
        month = jul,
          eid = {arXiv:2307.12546},
        pages = {arXiv:2307.12546},
          doi = {10.48550/arXiv.2307.12546},
archivePrefix = {arXiv},
       eprint = {2307.12546},
 primaryClass = {astro-ph.HE},
       adsurl = {https://ui.adsabs.harvard.edu/abs/2023arXiv230712546B}
}

@ARTICLE{2022A&C....4000609D,
       author = {{de Menezes}, R.},
        title = "{easyFermi: A graphical interface for performing Fermi-LAT data analyses}",
      journal = {Astronomy and Computing},
         year = 2022,
        month = jul,
       volume = {40},
          eid = {100609},
        pages = {100609},
          doi = {10.1016/j.ascom.2022.100609},
archivePrefix = {arXiv},
       eprint = {2206.11272},
 primaryClass = {astro-ph.HE},
       adsurl = {https://ui.adsabs.harvard.edu/abs/2022A&C....4000609D}
}

@ARTICLE{Shimwell2022,
       author = {{Shimwell}, T.~W. and {Hardcastle}, M.~J. and {Tasse}, C. and {Best}, P.~N. and {R{\"o}ttgering}, H.~J.~A. and {Williams}, W.~L. and {Botteon}, A. and {Drabent}, A. and {Mechev}, A. and {Shulevski}, A. and {van Weeren}, R.~J. and {Bester}, L. and {Br{\"u}ggen}, M. and {Brunetti}, G. and {Callingham}, J.~R. and {Chy{\.z}y}, K.~T. and {Conway}, J.~E. and {Dijkema}, T.~J. and {Duncan}, K. and {de Gasperin}, F. and {Hale}, C.~L. and {Haverkorn}, M. and {Hugo}, B. and {Jackson}, N. and {Mevius}, M. and {Miley}, G.~K. and {Morabito}, L.~K. and {Morganti}, R. and {Offringa}, A. and {Oonk}, J.~B.~R. and {Rafferty}, D. and {Sabater}, J. and {Smith}, D.~J.~B. and {Schwarz}, D.~J. and {Smirnov}, O. and {O'Sullivan}, S.~P. and {Vedantham}, H. and {White}, G.~J. and {Albert}, J.~G. and {Alegre}, L. and {Asabere}, B. and {Bacon}, D.~J. and {Bonafede}, A. and {Bonnassieux}, E. and {Brienza}, M. and {Bilicki}, M. and {Bonato}, M. and {Calistro Rivera}, G. and {Cassano}, R. and {Cochrane}, R. and {Croston}, J.~H. and {Cuciti}, V. and {Dallacasa}, D. and {Danezi}, A. and {Dettmar}, R.~J. and {Di Gennaro}, G. and {Edler}, H.~W. and {En{\ss}lin}, T.~A. and {Emig}, K.~L. and {Franzen}, T.~M.~O. and {Garc{\'\i}a-Vergara}, C. and {Grange}, Y.~G. and {G{\"u}rkan}, G. and {Hajduk}, M. and {Heald}, G. and {Heesen}, V. and {Hoang}, D.~N. and {Hoeft}, M. and {Horellou}, C. and {Iacobelli}, M. and {Jamrozy}, M. and {Jeli{\'c}}, V. and {Kondapally}, R. and {Kukreti}, P. and {Kunert-Bajraszewska}, M. and {Magliocchetti}, M. and {Mahatma}, V. and {Ma{\l}ek}, K. and {Mandal}, S. and {Massaro}, F. and {Meyer-Zhao}, Z. and {Mingo}, B. and {Mostert}, R.~I.~J. and {Nair}, D.~G. and {Nakoneczny}, S.~J. and {Nikiel-Wroczy{\'n}ski}, B. and {Orr{\'u}}, E. and {Pajdosz-{\'S}mierciak}, U. and {Pasini}, T. and {Prandoni}, I. and {van Piggelen}, H.~E. and {Rajpurohit}, K. and {Retana-Montenegro}, E. and {Riseley}, C.~J. and {Rowlinson}, A. and {Saxena}, A. and {Schrijvers}, C. and {Sweijen}, F. and {Siewert}, T.~M. and {Timmerman}, R. and {Vaccari}, M. and {Vink}, J. and {West}, J.~L. and {Wo{\l}owska}, A. and {Zhang}, X. and {Zheng}, J.},
        title = "{The LOFAR Two-metre Sky Survey. V. Second data release}",
      journal = {\aap},
         year = 2022,
        month = mar,
       volume = {659},
          eid = {A1},
        pages = {A1},
          doi = {10.1051/0004-6361/202142484},
archivePrefix = {arXiv},
       eprint = {2202.11733},
 primaryClass = {astro-ph.GA},
       adsurl = {https://ui.adsabs.harvard.edu/abs/2022A&A...659A...1S}
}

@ARTICLE{Manna2026,
       author = {{Manna}, Siddhant and {Desai}, Shantanu},
        title = "{A MINOT-based Study of Gamma-ray emission from SPT-CL J2012-5649/Abell 3667}",
      journal = {arXiv e-prints},
         year = 2026,
        month = may,
          eid = {arXiv:2605.20779},
        pages = {arXiv:2605.20779},
          doi = {10.48550/arXiv.2605.20779},
archivePrefix = {arXiv},
       eprint = {2605.20779},
 primaryClass = {astro-ph.HE},
       adsurl = {https://ui.adsabs.harvard.edu/abs/2026arXiv260520779M}
}

@ARTICLE{Planck2016,
       author = {{Planck Collaboration} and {Ade}, P.~A.~R. and {Aghanim}, N. and {Arnaud}, M. and {Ashdown}, M. and {Aumont}, J. and {Baccigalupi}, C. and {Banday}, A.~J. and {Barreiro}, R.~B. and {Barrena}, R. and {Bartlett}, J.~G. and {Bartolo}, N. and {Battaner}, E. and {Battye}, R. and {Benabed}, K. and {Beno{\^\i}t}, A. and {Benoit-L{\'e}vy}, A. and {Bernard}, J.-P. and {Bersanelli}, M. and {Bielewicz}, P. and {Bikmaev}, I. and {B{\"o}hringer}, H. and {Bonaldi}, A. and {Bonavera}, L. and {Bond}, J.~R. and {Borrill}, J. and {Bouchet}, F.~R. and {Bucher}, M. and {Burenin}, R. and {Burigana}, C. and {Butler}, R.~C. and {Calabrese}, E. and {Cardoso}, J.-F. and {Carvalho}, P. and {Catalano}, A. and {Challinor}, A. and {Chamballu}, A. and {Chary}, R.-R. and {Chiang}, H.~C. and {Chon}, G. and {Christensen}, P.~R. and {Clements}, D.~L. and {Colombi}, S. and {Colombo}, L.~P.~L. and {Combet}, C. and {Comis}, B. and {Couchot}, F. and {Coulais}, A. and {Crill}, B.~P. and {Curto}, A. and {Cuttaia}, F. and {Dahle}, H. and {Danese}, L. and {Davies}, R.~D. and {Davis}, R.~J. and {de Bernardis}, P. and {de Rosa}, A. and {de Zotti}, G. and {Delabrouille}, J. and {D{\'e}sert}, F.-X. and {Dickinson}, C. and {Diego}, J.~M. and {Dolag}, K. and {Dole}, H. and {Donzelli}, S. and {Dor{\'e}}, O. and {Douspis}, M. and {Ducout}, A. and {Dupac}, X. and {Efstathiou}, G. and {Eisenhardt}, P.~R.~M. and {Elsner}, F. and {En{\ss}lin}, T.~A. and {Eriksen}, H.~K. and {Falgarone}, E. and {Fergusson}, J. and {Feroz}, F. and {Ferragamo}, A. and {Finelli}, F. and {Forni}, O. and {Frailis}, M. and {Fraisse}, A.~A. and {Franceschi}, E. and {Frejsel}, A. and {Galeotta}, S. and {Galli}, S. and {Ganga}, K. and {G{\'e}nova-Santos}, R.~T. and {Giard}, M. and {Giraud-H{\'e}raud}, Y. and {Gjerl{\o}w}, E. and {Gonz{\'a}lez-Nuevo}, J. and {G{\'o}rski}, K.~M. and {Grainge}, K.~J.~B. and {Gratton}, S. and {Gregorio}, A. and {Gruppuso}, A. and {Gudmundsson}, J.~E. and {Hansen}, F.~K. and {Hanson}, D. and {Harrison}, D.~L. and {Hempel}, A. and {Henrot-Versill{\'e}}, S. and {Hern{\'a}ndez-Monteagudo}, C. and {Herranz}, D. and {Hildebrandt}, S.~R. and {Hivon}, E. and {Hobson}, M. and {Holmes}, W.~A. and {Hornstrup}, A. and {Hovest}, W. and {Huffenberger}, K.~M. and {Hurier}, G. and {Jaffe}, A.~H. and {Jaffe}, T.~R. and {Jin}, T. and {Jones}, W.~C. and {Juvela}, M. and {Keih{\"a}nen}, E. and {Keskitalo}, R. and {Khamitov}, I. and {Kisner}, T.~S. and {Kneissl}, R. and {Knoche}, J. and {Kunz}, M. and {Kurki-Suonio}, H. and {Lagache}, G. and {Lamarre}, J.-M. and {Lasenby}, A. and {Lattanzi}, M. and {Lawrence}, C.~R. and {Leonardi}, R. and {Lesgourgues}, J. and {Levrier}, F. and {Liguori}, M. and {Lilje}, P.~B. and {Linden-V{\o}rnle}, M. and {L{\'o}pez-Caniego}, M. and {Lubin}, P.~M. and {Mac{\'\i}as-P{\'e}rez}, J.~F. and {Maggio}, G. and {Maino}, D. and {Mak}, D.~S.~Y. and {Mandolesi}, N. and {Mangilli}, A. and {Martin}, P.~G. and {Mart{\'\i}nez-Gonz{\'a}lez}, E. and {Masi}, S. and {Matarrese}, S. and {Mazzotta}, P. and {McGehee}, P. and {Mei}, S. and {Melchiorri}, A. and {Melin}, J.-B. and {Mendes}, L. and {Mennella}, A. and {Migliaccio}, M. and {Mitra}, S. and {Miville-Desch{\^e}nes}, M.-A. and {Moneti}, A. and {Montier}, L. and {Morgante}, G. and {Mortlock}, D. and {Moss}, A. and {Munshi}, D. and {Murphy}, J.~A. and {Naselsky}, P. and {Nastasi}, A. and {Nati}, F. and {Natoli}, P. and {Netterfield}, C.~B. and {N{\o}rgaard-Nielsen}, H.~U. and {Noviello}, F. and {Novikov}, D. and {Novikov}, I. and {Olamaie}, M. and {Oxborrow}, C.~A. and {Paci}, F. and {Pagano}, L. and {Pajot}, F. and {Paoletti}, D. and {Pasian}, F. and {Patanchon}, G. and {Pearson}, T.~J. and {Perdereau}, O. and {Perotto}, L. and {Perrott}, Y.~C. and {Perrotta}, F. and {Pettorino}, V. and {Piacentini}, F. and {Piat}, M. and {Pierpaoli}, E. and {Pietrobon}, D. and {Plaszczynski}, S. and {Pointecouteau}, E. and {Polenta}, G. and {Pratt}, G.~W. and {Pr{\'e}zeau}, G. and {Prunet}, S. and {Puget}, J.-L.},
        title = "{Planck 2015 results. XXVII. The second Planck catalogue of Sunyaev-Zeldovich sources}",
      journal = {\aap},
         year = 2016,
        month = sep,
       volume = {594},
          eid = {A27},
        pages = {A27},
          doi = {10.1051/0004-6361/201525823},
archivePrefix = {arXiv},
       eprint = {1502.01598},
 primaryClass = {astro-ph.CO},
       adsurl = {https://ui.adsabs.harvard.edu/abs/2016A&A...594A..27P}
}

@INPROCEEDINGS{Wood2017,
       author = {{Wood}, M. and {Caputo}, R. and {Charles}, E. and {Di Mauro}, M. and {Magill}, J. and {Perkins}, J.~S. and {Fermi-LAT Collaboration}},
        title = "{Fermipy: An open-source Python package for analysis of Fermi-LAT Data}",
    booktitle = {35th International Cosmic Ray Conference (ICRC2017)},
         year = 2017,
       series = {International Cosmic Ray Conference},
       volume = {301},
        month = jul,
          eid = {824},
        pages = {824},
          doi = {10.22323/1.301.0824},
archivePrefix = {arXiv},
       eprint = {1707.09551},
 primaryClass = {astro-ph.IM},
       adsurl = {https://ui.adsabs.harvard.edu/abs/2017ICRC...35..824W}
}
\appendix

\end{document}